\documentclass{article}
\usepackage[utf8]{inputenc}

\usepackage{arxiv}

\usepackage[T1]{fontenc}    
\usepackage{hyperref}       
\usepackage{url}            
\usepackage{booktabs}       
\usepackage{amsfonts}       
\usepackage{nicefrac}       
\usepackage{microtype}      
\usepackage{pdflscape}

\usepackage{amssymb}
\usepackage{commath}
\usepackage{multirow}
\usepackage{tabu}
\usepackage[table]{xcolor}
\usepackage{mdframed}
\usepackage{placeins}
\usepackage{floatrow}
\usepackage[labelfont=bf]{caption}
\usepackage[caption=false]{subfig}

\usepackage{algorithm}
\usepackage[noend]{algpseudocode}

\usepackage{tikz}

\usepackage{graphicx}
\usepackage{pbox}

\usepackage[nameinlink,capitalize]{cleveref}
\usepackage{bm}

\usepackage[sorting=none,
            style=numeric-comp,
            bibstyle=ieee,
            url=false,
            doi=false,
            isbn=false,
            language=english,
            clearlang=true,
            maxbibnames=5,
            arxiv=false,
            giveninits=true]{biblatex}
            
\usepackage{biblatex-shortfields}

\DeclareFieldFormat{ieeecase}{#1}
\DeclareNameAlias{sortname}{first-last}

\renewbibmacro{in:}{}

\DeclareFieldFormat[article]
{volume}{\textbf{#1}} 
\DeclareFieldFormat[article, inbook, incollection, inproceedings, misc, thesis, unpublished]
{number}{(#1)}
\DeclareFieldFormat[article,inproceedings]
{pages}{#1}
\DeclareFieldFormat[inbook, incollection]
{pages}{p. #1}
\renewbibmacro*{volume+number+eid}{
  \printfield{volume}\printfield{number}:
}

\AtEveryBibitem{\clearfield{month}}
\AtEveryBibitem{\clearfield{day}}
\AtEveryBibitem{%
  \clearlist{language}%
}
 
\bibliography{references}

\usepackage{mathtools}

\usepackage{mhchem}

\newcommand{\singlereac}[1]{\xrightarrow{\hspace{1ex}#1\hspace{1ex}}}
\newcommand{\doublereac}[2]{\ce{<=>[{#1}][{#2}]}}

\newcommand{\N}{\mathbb{N}}
\newcommand{\R}{\mathbb{R}}

\newcommand{\Wass}{\mathcal W}

\newcommand{\E}{\mathbb{E}}

\DeclareMathOperator{\GP}{GP}

\renewcommand{\dif}{\mathrm{d}}

\newcommand{\bvec}[1]{\vec{\boldsymbol{#1}}}
\def\eps{\epsilon}

\usepackage{layouts}

\newcommand{\bx}{{\bf x}}

\def\pltlinewidth{0.8pt}

\definecolor{col1}{HTML}{4da9ff}
\definecolor{col2}{HTML}{FFB300}
\definecolor{col3}{HTML}{00b300}

\newcommand{\drawsolidbox}[1]{{\tikz \draw[#1!40!white,fill=#1!40!white,line width=0pt] (0,0) rectangle (0.19,0.19) ; }}

\newcommand{\drawsolidboxtwo}[1]{{\tikz \draw[#1!60!white,fill=#1!60!white,line width=0pt] (0,0) rectangle (0.19,0.19) ; }}

\DeclareRobustCommand{\drawbox}[1]{{\tikz \draw[#1,fill=#1!40!white,line width=1pt] (0,0) rectangle (0.18,0.18) ; }}
\DeclareRobustCommand{\drawsmallbox}[1]{{\tikz \draw[#1,fill=#1!40!white,line width=1pt] (0,0) rectangle (0.12,0.12) ; }}

\newcommand{\rulesep}{\unskip\ \vrule\ }

\title{Parameter estimation for biochemical reaction networks using Wasserstein distances}

\author{
  Kaan \"Ocal \\
  School of Informatics \\ 
  University of Edinburgh \\ 
  Edinburgh EH8 9AB \\
  United Kingdom \\
  \texttt{kaan.ocal@ed.ac.uk}\\
   \And
  Ramon Grima \\
  School of Biological Sciences \\
  University of Edinburgh \\
  Edinburgh EH9 3JH \\
  United Kingdom \\
  \texttt{ramon.grima@ed.ac.uk} \\
   \And
  Guido Sanguinetti \\
  School of Informatics \\ 
  University of Edinburgh \\ 
  Edinburgh EH8 9AB \\
  United Kingdom \\
  \texttt{gsanguin@inf.ed.ac.uk} \\
}

\begin{document}
\maketitle

\begin{abstract}
    We present a method for estimating parameters in stochastic models of biochemical reaction networks by fitting steady-state distributions using Wasserstein distances. We simulate a reaction network at different parameter settings and train a Gaussian process to learn the Wasserstein distance between observations and the simulator output for all parameters. We then use Bayesian optimization to find parameters minimizing this distance based on the trained Gaussian process. The effectiveness of our method is demonstrated on the three-stage model of gene expression and a genetic feedback loop for which moment-based methods are known to perform poorly. Our method is applicable to any simulator model of stochastic reaction networks, including Brownian Dynamics.
\end{abstract}

\keywords{Wasserstein distance \and Bayesian optimization \and Chemical Master Equation \and parameter estimation}

\section{Introduction}

Modern experimental methods such as flow cytometry and fluorescence in-situ hybridization (FISH) allow the measurement of cell-by-cell molecule numbers for RNA, proteins and other substances for large numbers of cells at a time, opening up new possibilities for the quantitative analysis of biological systems. Of particular interest is the study of biological reaction systems describing processes such as gene expression, cellular signalling and metabolism on a molecular level. It is well established that many of these processes are inherently stochastic \cite{elowitz_stochastic_2002,choi_stochastic_2008,kiviet_stochasticity_2014} and that deterministic approaches can fail to capture properties essential for our understanding of these systems \cite{mcadams_its_1999,ramaswamy_discreteness-induced_2012}. Despite recent technological and conceptual advances, modelling and inference for stochastic models of reaction networks remains challenging due to additional complexities not present in the deterministic case. The Chemical Master Equation (CME) \cite{van_kampen_stochastic_2007} in particular, while frequently used to model many types of reaction networks, is difficult to solve exactly, and parameter inference in practice often relies on a variety of approximation schemes whose accuracy and efficiency can vary widely and unpredictably depending on the context \cite{schnoerr_approximation_2017}. 

The diffusion approximation and the system size expansion \cite{van_kampen_stochastic_2007} are two well-known approximations to the CME and frequently used for inference \cite{golightly_bayesian_2006,golightly_bayesian_2011}. However, in the presence of bimolecular reactions such as enzyme-substrate or protein-DNA interactions these approximations are unable to deal with low copy numbers frequently found in biological systems, rendering them unsuitable for inference in reaction systems involving species such as e.g.~individual genes or mRNA, which is often present in copy numbers of less than $20$ per cell \cite{marguerat_quantitative_2012}.

Other methods for parameter inference rely on fitting moments of the particle number distributions returned by the CME to experimental data \cite{frohlich_inference_2016, zechner_moment-based_2012,ruess_moment-based_2015,cinquemani_identifiability_2018}. Moments can often be computed or approximated by solving a set of coupled equations, bypassing expensive simulations of the system in question. However, as pointed out in \cite{ruess_moment-based_2015,neuert_systematic_2013,munsky_distribution_2018} these moment-based methods are not always suitable for inference. Computing the moments for reaction systems with bimolecular interactions usually necessitates the use of so-called moment closure approximations, validity conditions for which are not well-understood \cite{schnoerr_comparison_2015,schnoerr_validity_2014,schilling_adaptive_2016}. Given the wide variety of moment closure schemes it is not generally clear \emph{a priori} which, if any, will prove suitable for a given reaction system, and the right method is usually chosen empirically based on its performance \cite{cao_accuracy_2019}. In addition, moment closure typically results in a set of coupled nonlinear equations which can have multiple different solutions, further complicating their use in parameter inference. We will provide an example of a genetic feedback loop based on \cite{cao_accuracy_2019} for which many commonly used moment approximations break down or provide inaccurate moment estimates. 

In this paper we propose a method to estimate parameters for the Chemical Master Equation from population snapshot data by matching steady state distributions using Wasserstein distances \cite{villani_optimal_2009}, also known as Earth Mover's distances in the literature. Since Wasserstein distances can in general not be computed analytically we take sample-based estimates and emulate the complex dependency of the output on the parameters by using a Bayesian regression approach. We train a Gaussian process (GP) to learn the distances between the observed data and the steady-state distributions at different parameter settings, obtained using simulations, and apply Bayesian optimization (BO) to find the parameters minimizing this distance, sequentially selecting the next parameter settings to evaluate until the optimum is found. Our approach requires orders of magnitude fewer evaluations than grid searches and is suitable for any simulator-based model of reaction networks, including models such as Brownian Dynamics for which the previously mentioned inference methods are not available.

The idea of performing parameter inference based on considering full distributions has been previously explored in \cite{neuert_systematic_2013} and \cite{munsky_distribution_2018}, where it is shown that estimating parameters by matching moments can result in reduced predictive power and inaccurate fits of the actual distributions. Both \cite{neuert_systematic_2013} and  \cite{munsky_distribution_2018} perform parameter inference by maximizing the likelihood of the observed data; this requires computing the likelihood function which is commonly done by solving a finite-dimensional approximation of the Chemical Master Equation, the so-called Finite State Projection (FSP) \cite{munsky_finite_2006}. The FSP involves solving a large system of coupled ordinary differential equations and scales poorly for larger reaction systems as the number of equations grows exponentially with the number of species considered. Our method in contrast does not require the likelihood function and instead relies on empirically approximating the steady state distribution using simulations, rendering it more scalable and more flexible than current likelihood-based approaches.

One drawback of the method we present is that is only provides point estimates for parameters and does not return a measure of confidence that would enable one to assess uncertainty. While Bayesian optimization is used to infer the location of the optimal parameters it does not yield a posterior distribution over parameters as in Bayesian inference. To the knowledge of the authors efficient Bayesian inference for the CME based on steady-state observation data remains an intractable problem in general.

\section{Background}
\subsection{The Chemical Master Equation}

In this section we briefly review biochemical reaction networks and the Chemical Master Equation, referring to \cite{schnoerr_approximation_2017} for a more comprehensive treatment. A reaction network consists of species $S_j$, $j=1,\ldots,s$ and reactions $R_i$, $i=1,\ldots,r$ of the form
\begin{align}
    a_{i,1} S_1 + \ldots + a_{i,s} S_s \longrightarrow b_{i,1} S_1 + \ldots + b_{i,s} S_s
\end{align}

\noindent where $\bvec a_i := (a_{i,1},\ldots,a_{i,s})$ and $\bvec b_i := (b_{i,1},\ldots,b_{i,s})$ are vectors of nonnegative integers, the stoichiometric coefficients of the reaction. The Chemical Master Equation approach models the reaction network as a Markov chain whose states are given by tuples $\bvec n := (n_1,\ldots,n_s) \in \N^s$ defining the number of particles of each species present at each time. The transitions of the Markov chain correspond to reactions, with the transition rate of reaction $R_i$ determined by the state-dependent propensity function $\rho_i(\bvec n)$. The forward Kolmogorov equation for this Markov chain is called the Chemical Master Equation and reads:
\begin{align}
    \frac{\partial}{\partial t} P(\bvec n,t) &= \sum_{i=1}^r \left[\rho_i\left(\bvec n - \bvec S_i\right) P\left(\bvec n - \bvec S_i,t\right) - \rho_i(\bvec n) P(\bvec n,t) \right] \label{eq:cme}
\end{align}

\noindent Here $\bvec S_i := \bvec b_i - \bvec a_i$ describes the net change in reactant numbers during reaction $i$.

The form of the transition functions $\rho_j(\vec n)$ depends on the specific reaction, but the most commonly used transition functions are given by the mass-action law,
\begin{align}
    \rho_i(\bvec n) := \lambda_i \binom{n_1}{a_{i,1}} \ldots \binom{n_s}{a_{i,s}} \label{eq:mass_action}
\end{align}

\noindent for rate constants $\lambda_i > 0$. While our approach can handle general transition functions, in what follows we restrict ourselves to mass-action propensities of the form \labelcref{eq:mass_action}. With this setup the task of inferring parameters for the CME reduces to finding the appropriate rate constants $\lambda_i$.

We remark that the steady state distribution of a reaction system does not change if all transition rates are rescaled by a common factor $c > 0$. Thus by observing the steady state one can only identify the rate constants up to a common scaling factor, which can be fixed if any one reaction rate is known. It is frequently possible to measure the degradation rate of reaction species experimentally, which removes this ambiguity - such an approach to inference is taken for example in \cite{schwanhausser_global_2011}. In the remainder of this paper we will always assume that one reaction rate is given and estimate the remaining rate constants.

\subsection{Wasserstein distances}

\label{sec:wasserstein}

We perform parameter estimation based on population snapshot data by minimizing the discrepancy between the observed distribution over particle numbers and the distributions returned by the simulator. In this section we motivate and describe our choice of discrepancy measure, namely Wasserstein distances and refer to \cite{ashyraliyev_systems_2009} for a general discussion of parameter estimation in biology. 

\begin{figure}[!tb]
    \begin{flushright}
    \small
    \vspace{0.2cm}
    \begin{tabu}{|[\pltlinewidth]l|[\pltlinewidth]c c c c @{\hskip 1.5ex} c @{\hskip 1ex} c|[\pltlinewidth]c}
        \tabucline[\pltlinewidth]{2-7} 
        \multicolumn{1}{l|[\pltlinewidth]}{} & \textbf{1-Wass} & \textbf{TV} & \textbf{Hell} & \textbf{Jen} & {\footnotesize $\bm{\mathrm{KL}(1 \, \| \, 2)}$} & {\footnotesize $\bm{\mathrm{KL}(2 \, \| \, 1)}$} & \hspace{0.3cm} \\        
        \tabucline[\pltlinewidth]{1-7}
        $d(\drawbox{col1},\drawbox{col2})$ & 14.8 & 0.999 & 0.997 & $\infty$ & $\infty$ & $\infty$ \\
        $d(\drawbox{col1},\drawbox{col3})$ & 48.9 & 1.000 & 1.000 & $\infty$ & $\infty$ & $\infty$ \\ 
        $d(\drawbox{col2},\drawbox{col3})$ & 34.0 & 0.999 & 0.996 & $\infty$ & $\infty$ & $\infty$
        \\
        \tabucline[\pltlinewidth]{1-7}
    \end{tabu}
    \end{flushright}
    
    \vspace{-2.2cm}
    
    \includegraphics{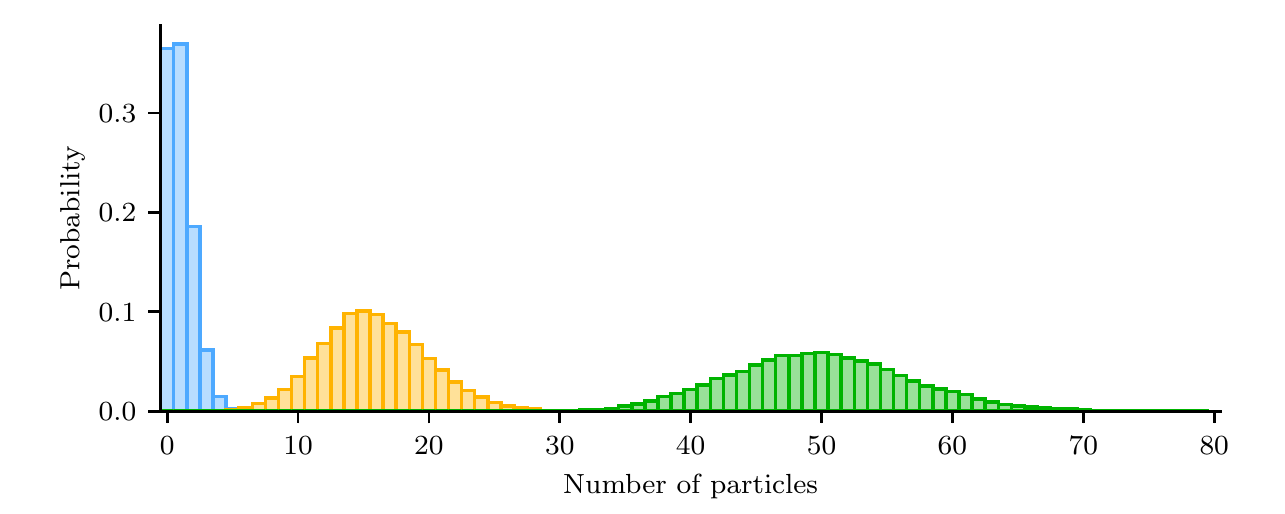}

    \centering
    
    \caption{Common discrepancy measures for probability distributions do not provide usable distance metrics for simulated data. The histograms show empirical estimates of the steady-state distributions for a simple birth-death process with three different ratios of birth and death rates $r$ ($r_{\drawsmallbox{col1}} = 1$, $r_{\drawsmallbox{col2}} = 16$, $r_{\drawsmallbox{col3}} = 50$). The table compares their $1$-Wasserstein ($1$-Wass), total variation (TV) and Hellinger (Hell) distances, their Jensen divergences (Jen) and their Kullback-Leibler divergences (KL). Even though the two outer histograms are significantly further apart than the neighbouring pairs, the total variation and Hellinger distances in all cases differ by less than $1\%$, and the Kullback-Leibler and Jensen divergences between any two of these histograms are infinite. The \mbox{$1$-Wasserstein} distance on the other hand captures an intuitive notion of distance between these histograms.}
    \label{fig:dist_TV_rebuttal}
\end{figure}

There are a variety of commonly used discrepancy measures in the literature that can be used to compare and match distributions directly. Information-theoretic measures like the Kullback-Leibler or Jensen divergences often become infinite if the compared distributions do not have identical support (which is rarely the case for the empirical distributions we consider), rendering them essentially useless for our purposes. Other metrics like the total variation and Hellinger distances do not provide a usable measure of distance for distributions without significant overlap, a common scenario where particle numbers can vary over orders of magnitude (see \cref{fig:dist_TV_rebuttal}). In contrast, Wasserstein distances are generally well-defined and provide an interpretable distance metric between distributions. This is important for our global approach used for optimization described in \cref{sec:bayopt}.
 
\begin{figure}[!hbt]
    \centering
    
    \begin{minipage}[t]{0.4\textwidth}
    \sidesubfloat[]{
    \hspace{-0.8cm}
    \includegraphics{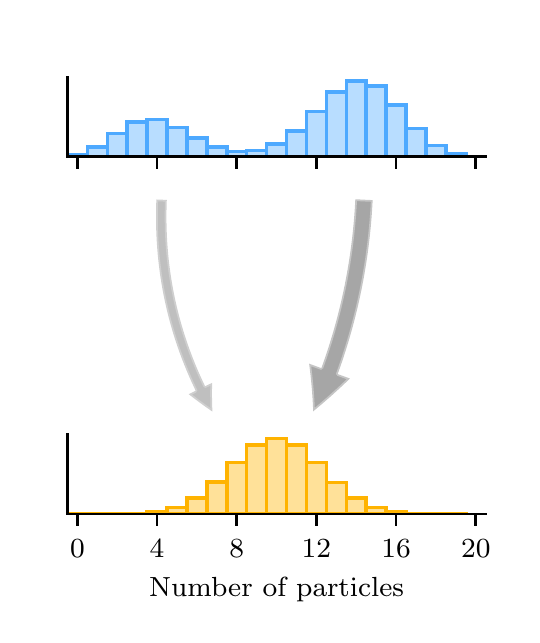}
    
    
    }
    \end{minipage}
    ~
    \rulesep
    ~
    \begin{minipage}[t]{0.55\textwidth}
    \sidesubfloat[]{
    \hspace{-0.8cm}
    \includegraphics{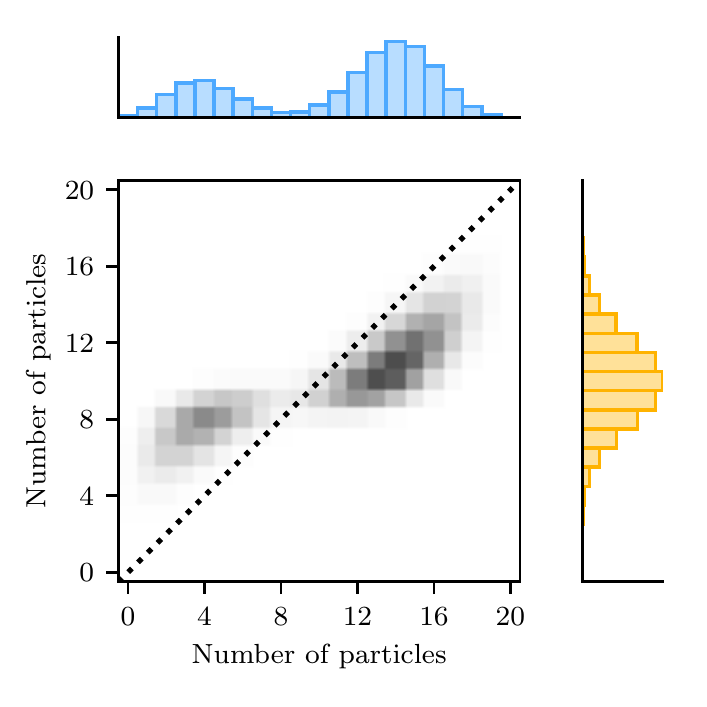}
    }
    \end{minipage}
    
    \caption{\textbf{(a)} Optimal transport distances between histograms measure how much mass has to be moved in order to convert one histogram into the other. \textbf{(b)} Illustration of a transport plan between the two histograms in (a). The joint histogram shows the amount of mass transported between different locations in the histograms. Mass on the diagonal (dotted line) is not moved during transport.}
    \label{fig:dist_ot}
\end{figure}

The steady-state distribution for a reaction system can be represented as a normalized histogram over the state space $\N^s$, where $s$ is the number of reactant species. Wasserstein distances measure the discrepancy between two such histograms by considering how much and how far probability mass has to be moved in order to reshape one histogram into the other; for this reason they are often called Earth Mover's Distances (see \cref{fig:dist_ot}). The concept of moving mass between histograms is formalized by transport plans introduced below.

Consider two histograms $P$, $Q$ over $\N^s$; the value of the histogram $P$ at $\vec i = (i_1,\ldots,i_s) \in \N^s$ is denoted $P_{\vec i}$. A transport plan $T$ between $P$ and $Q$ is a histogram on $\N^s \times \N^s$ whose first and second marginals are $P$ and $Q$, respectively,
\begin{align}
    \sum_{\vec j} T_{\vec i, \vec j} &= P_{\vec i} & \sum_{\vec i} T_{\vec i, \vec j} &= Q_{\vec j}  \label{eq:ot_coupling_def}
\end{align}

\noindent The value $T_{\vec i,\vec j}$ can be viewed as the amount of probability mass that has to be moved from $\vec i$ to $\vec j$ in order to convert the histogram $P$ into $Q$; \cref{eq:ot_coupling_def} then represents the conservation of probability mass during this process. The simplest transport plan between $P$ and $Q$ is the independent coupling given by
\begin{align}
    (P \otimes Q)_{\vec i, \vec j} &= P_{\vec i} \cdot Q_{\vec j}
\end{align}

\noindent which specifies that the probability mass in every bin of $P$ is to be distributed evenly across $Q$. We denote the space of transport plans between $P$ and $Q$ by $U(P,Q)$. 

Optimal transport maps are defined by assigning to each move a certain cost. We define a cost function $C$ to be a nonnegative function on $\N^s \times \N^s$, where $C_{\vec i, \vec j}$ represents the cost involved in transporting a unit of probability mass from $\vec i$ to $\vec j$. While this is not necessary for the theory we will assume that the cost function is a distance metric on the ground space $\N^s$, ie. there is a metric $d$ on $\N^s$ such that $C_{\vec i, \vec j} = d(\vec i, \vec j)$. The optimal transport problem with cost function $C$ reads
\begin{align}
    \mathcal W_C(P, Q) := \inf_{T \in U(P,Q)} \langle C, T \rangle &= \inf \left\{ \sum_{\vec i, \vec j} C_{\vec i, \vec j} T_{\vec i,\vec j} \colon T_{\vec i, \vec j} \geq 0, \right. \nonumber \\
    &\left. \qquad \qquad \sum_{\vec j} T_{\vec i, \vec j} = P_{\vec i},  \sum_{\vec i} T_{\vec i, \vec j} = Q_{\vec j} \right\} \label{eq:ot_wasserstein_concrete}
\end{align}

\noindent One can check that the Wasserstein distance $\mathcal W_C$ defines a metric on the space of probability distributions on $\N^s$. More generally one can verify that the $p$-Wasserstein distance
\begin{align}
    \mathcal W_C^{(p)}(P, Q) :=  \mathcal W_{C^p}(P, Q)^{1/p}
\end{align}

\noindent defines a metric on the space of probability distributions on $\N^s$ for all $p \geq 1$. Here the cost is given by $C^p_{\vec i, \vec j} = d(\vec i, \vec j)^p$.

A commonly used class of metrics on the space $\N^s$ is given by the weighted $\ell^q$-metrics defined by
\begin{align}
    d_{\ell^q}^{(\bf w)}(\vec i, \vec j) &= \left(\sum_{k=1}^s w_k^q \abs{i_k - j_k}^q \right)^{\frac 1 q}
\end{align}

\noindent for $q \geq 1$ and positive weights $w_k$. These metrics are all equivalent and yield equivalent classes of $p$-Wasserstein metrics for fixed $p$; the exact choice of $q$ and the weights is therefore not important in most applications. We will use the notation $\mathcal W_p(P,Q)$ for the $p$-Wasserstein distance when the ground metric is understood.
\subsection{Bayesian optimization}

\label{sec:bayopt}

Since the dependence of the Wasserstein distance between observed data and simulator output on the parameters of the CME is not available in closed form and can only be evaluated by running simulations we are faced with the task of minimizing a function that is expensive to evaluate and about which no gradient information is available. In order to do this efficiently we rely on Bayesian optimization, a method for efficiently optimizing expensive black-box functions in low to moderate dimensions based on a Gaussian process surrogate of the function to be optimized. See \cite{rasmussen_gaussian_2006} for a comprehensive reference on Gaussian processes and \cite{shahriari_taking_2016} for an overview of Bayesian optimization going beyond the description in this paper. 

In order to apply Bayesian optimization to our problem we start by placing a Gaussian process prior on the loss function $L(\bx)$, which is defined on a space $\mathcal X$, that measures the discrepancy with the observed data:
\begin{align}
    \hat L \sim \GP(\mu(\bx), k(\bx,\bx'))
\end{align}

\noindent with mean function $\mu(\bx)$ and covariance kernel $k(\bx,\bx')$. Thus $\hat L$ is a statistical model of the true function $L$. In our case $L$ will be the Wasserstein loss measuring the Wasserstein distance between the steady-state distributions, $\mathcal X$ will be the chosen space of parameters and the $\bx$ will be individual parameter settings. We assume that we can use a simulator to compute noisy observations of $L(\bx_i)$:
\begin{align}
    \tilde L(\bx_i) = L(\bx_i) + \eps_i \label{eq:obs_noise}
\end{align}

\noindent at any given point $\bvec x_i$, where the $\eps_i$ are observation noise. We assume that the $\eps_i$ are iid.~normal random variables with mean zero; the standard deviation of the observation noise can be interpreted as a hyperparameter of the Gaussian process. With this setup our Gaussian process $\hat L$ can be updated by obtaining data points $\mathcal D_i = \{ \bx_i, \tilde L(\bx_i) \}$ for different $\bx_i$ and computing the posterior $\hat L \; | \; \mathcal D$. 

Our goal is to minimize $L(\bx)$ with as few evaluations of $\tilde L(\bx)$ as possible. Bayesian optimization consists of a procedure for sequentially choosing the points $\bx_1, \ldots, \bx_n \in \mathcal X$ at which $\tilde L(\bx)$ is to be evaluated in order to decrease the uncertainty about the location of the optimum, based on the Gaussian process $\hat L$. This is done by considering a so-called acquisition function $\alpha(\bx; L \, | \, \mathcal D)$ depending on the collected observations $\mathcal D$ and choosing the next point to evaluate as
\begin{align}
    \bx_{n+1} = \arg \max_{\bx \in \mathcal X} \alpha(\bx; L \; | \; \mathcal D_{1:n})
\end{align}

\noindent The acquisition function returns a point $\bx_{n+1}$ such that computing $\tilde L(\bx_{n+1})$ yields additional knowledge about the minimum of $L(\bx)$, e.g. by choosing a point which is likely to be near the true minimum. It should be simpler to evaluate and optimize than the target function so that one can use standard optimization methods for finding $\bx_{n+1}$ with little overhead. After finding $\bx_{n+1}$ and running the simulator to compute $\tilde L(\bx_{n+1})$ one updates the Gaussian process $\hat L$ with the data $\mathcal D_{n+1} = \{ \bx_{n+1}, \tilde L(\bx_{n+1}) \}$ and repeats this procedure until the true optimum of $L(\bx)$ is found. An illustration of Bayesian optimization can be seen in \cref{fig:BO_1D}.


A common choice for the acquisition function $\alpha$ is Expected Improvement, defined by the following formula:
\begin{align}
    \alpha_{\mathrm{EI}}(\bx; L \; | \; \mathcal D) := \E_{L | \mathcal D} \left[ \left(\min_{\bx_i \in \mathcal D} \tilde L(\bx_i) - \hat L(\bx) - \beta\right)^{+}\right] \label{eq:ei}
\end{align}

\noindent Here $\beta \geq 0$ is a small ``jitter'' parameter used to reduce the time spent in local optima and increase exploration. With this acquisition function the predicted optimum of $L(\bx)$ is typically computed as:
\begin{align}
    \bx^* := \arg \min_{\bx_i \in \mathcal D} \tilde L(\bx_i)
\end{align}

\noindent Since \cref{eq:ei} can be computed in closed form the expected improvement at a point $\bx$ can be evaluated quite cheaply, and gradients can be computed at little additional cost. It is known that Bayesian optimization using this acquisition function is guaranteed to find the optimum of the target function $L$ under some mild assumptions on $L$ and the Gaussian process prior \cite{vazquez_convergence_2010}. This combined with its simplicity and empirical performance properties make Expected Improvement a popular choice of acquisition function in Bayesian optimization. Other common acquisition functions are Upper Confidence Bound, Probability of Improvement and Knowledge Gradient \cite{shahriari_taking_2016}, which we shall not consider here.

\setcounter{figure}{0}
\begin{figure}[!h]
\floatbox[{\capbeside\thisfloatsetup{capbesideposition={right,center},capbesidewidth=0.45\textwidth}}]{figure}[\FBwidth]
{\caption{Illustration of Bayesian optimization in one dimension. Plotted are a Gaussian process (GP) and its acquisition function (AF) before and after an update step. The shaded area represents two standard deviations around the mean. Each round consists of computing the loss function at the point maximizing the acquisition function (vertical line) and updating the GP with the computed value. After the update step the acquisition function changes to reflect the information gained in the process, and a new point is chosen for the next round.}\label{fig:BO_1D}}
{
    \centering
    \begin{minipage}{0.5\textwidth}
    
    \subfloat[Before update]{
    \includegraphics{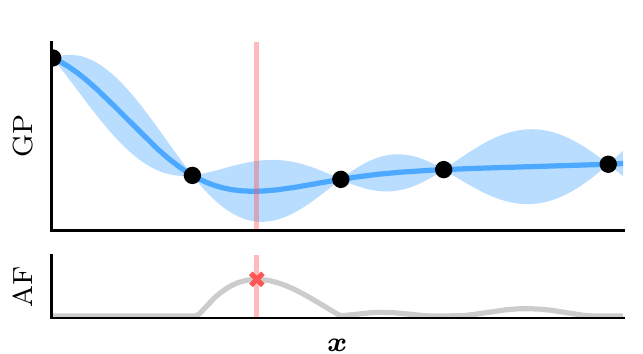}
    }
    
    \subfloat[After update]{
    \includegraphics{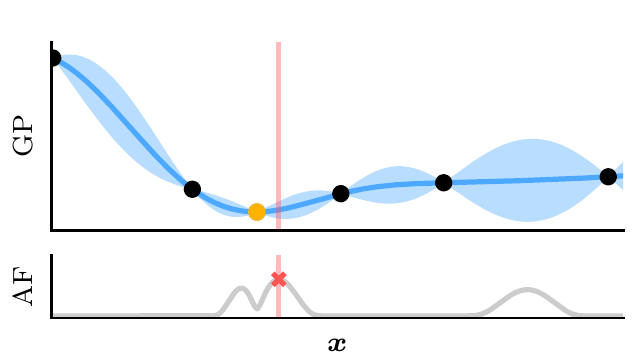}
    }
    \end{minipage}
}
\end{figure}

\section{Methods}

\subsection{Computing Wasserstein distances}

\label{sec:methods_wass}

Computing Wasserstein distances by finding an explicit optimal transport map in \cref{eq:ot_wasserstein_concrete} requires solving a constrained linear optimization problem with a large number of variables and constraints. As this approach quickly gets impracticable for realistic histogram sizes we use an adaptation of the Sinkhorn algorithm presented in \cite{cuturi_sinkhorn_2013}, which computes the optimum of a relaxed version of the transport problem:
\begin{align}
    \mathcal W^{(\eps)}_C(P,Q) = \inf_{T \in U(P,Q)} \langle C, T \rangle - \eps H(T) \label{eq:ot_sinkhorn}
\end{align}

\noindent where the regularizer $H(T)$ is defined as
\begin{align}
    H(T) = -\sum_{\vec i, \vec j} T_{\vec i, \vec j} \log T_{\vec i, \vec j}
\end{align}

\noindent with the convention that $H(T) = -\infty$ if one of the entries of $T$ is not positive. One can derive error bounds on the difference $\mathcal W^{(\eps)}_C(P,Q) - \mathcal W_C(P,Q)$ and show that the solutions to \cref{eq:ot_sinkhorn} converge to the solution of the unregularized problem as $\eps \rightarrow 0$. We can thus compute Wasserstein distances approximately by solving \cref{eq:ot_sinkhorn} for small enough $\eps$. 

The Sinkhorn algorithm is an iterative solver for \cref{eq:ot_sinkhorn}, a description of which can be found in \cite{cuturi_sinkhorn_2013}. One drawback of this algorithm is that the number of iterations required for convergence increases as $\eps \rightarrow 0$; in order to compute the solution to \cref{eq:ot_sinkhorn} for small $\eps$ we therefore use an annealing procedure starting with a large value of $\eps$ (typically $\eps = 10$) and multiplying it by an annealing factor $\delta < 1$ at each step. We run each step until the two margin constraints \labelcref{eq:ot_coupling_def} are satisfied to a specified tolerance $\eps'$ in the $\ell^1$-norm. To improve convergence speed for small $\eps$ we use the overrelaxation method presented in \cite{thibault_overrelaxed_2017}.

Due to the large numerical range encountered in the Sinkhorn algorithm all computations are performed in log-space. In addition, when computing the $p$-Wasserstein distance with a weighted $\ell^q$ ground metric, for $p = q$ the cost matrix $C_{\vec i,\vec j}^p = d_{\ell^q}^{(\bf w)}(\vec i, \vec j)^p$ decomposes as a sum of terms, one for each histogram dimension,
\begin{align}
    C_{\vec i,\vec j}^p = \sum_{k=1}^s w_k^p \abs{i_s - j_s}^p \label{eq:wass_cost_separable}
\end{align}

\noindent which permits vectorization of the relevant matrix-vector products in the Sinkhorn algorithm. For this reason and since the exact choice of $q$ does not matter (see \cref{sec:wasserstein}) we will generally set $p = q$ for efficiency. 

We remark that (unweighted) Wasserstein distances in one dimension can be evaluated using a simpler and more straightforward algorithm: if $F$ and $G$ are the cumulative distribution functions of two probability distributions $f$ and $g$ on $\R$, respectively, then one can prove \cite{villani_optimal_2009} that 
\begin{align}
    \Wass_p(f,g) &= \left( \int_0^1 \abs{F^{-1}(x) - G^{-1}(x)}^p \dif x \right)^{1/p}
\end{align}

\noindent For discrete histograms this integral can be computed exactly with little overhead, making Wasserstein distances in one dimension especially suited for computational purposes. 

\subsection{Bayesian optimization}

\label{sec:impl_bayopt}

At this stage we are given experimentally observed data $P$ (a histogram over particle numbers) and a parametrized simulator model $Q(\bx)$ whose output is a histogram depending on the parameters $\bx \in \mathcal X$. In our case the function $Q(\bx)$ is the steady state distribution of the system with parameters $\bx$. Our goal is to find parameters $\bx$ minimizing a chosen Wasserstein distance 
\begin{align}
    L(\bx) := \mathcal W_p(Q(\bx), P)
\end{align}

\noindent We can approximate $L$ for any set of parameters by simulating the Chemical Master Equation for sufficiently long times using the Stochastic Simulation Algorithm \cite{gillespie_general_1976}. As described in \cref{sec:bayopt} we optimize the loss using Bayesian optimization; in this section we describe the details of our setup.

Assuming our task is to infer $d$ different parameters we start by choosing a (bounded) search space $\mathcal X \subseteq \R^d$. In practice one should choose a reasonably large region in which the true parameters are expected to be found; if this is not the case after optimization one can enlarge the search space and continue optimization until the optimum is found. We sample $m$ points $\bx_1,\ldots,\bx_m \in \mathcal X$ spread across the search space and evaluate $\tilde L(\bx_1), \ldots, \tilde L(\bx_m)$ in order to pre-train the GP. The choice of $m$ usually depends on the dimension and the expected roughness of the loss landscape. We sample the points using Latin hypercubes in order to achieve uniform coverage of the parameter space. 

The mean of the Gaussian process $\hat L$ is set to a constant equal to the mean of the $\tilde L(\bx_i)$. For the covariance kernel we initially choose a squared exponential function of the form
\begin{align}
    k(\bx, \bx') = \sigma_y^2 \exp\left[- \frac 1 2 (\bx - \bx')^T \Lambda (\bx - \bx') \right]
\end{align}

\noindent The hyperparameters for this setup are the marginal variance $\sigma_y^2$ and the precision matrix $\Lambda$, restricted to be diagonal for simplicity. We fit the kernel hyperparameters by maximizing the marginal likelihood of the data $\mathcal D_{1:m}$, a common procedure for determining hyperparameters for Gaussian process regression \cite{rasmussen_gaussian_2006}.

Bayesian optimization now consists of repeatedly optimizing the acquisition function, computing the loss function $L(\bx^*)$ at the optimum $\bx^*$ and updating the Gaussian process $\hat L$ with this information. In order to improve the fit of the Gaussian process we periodically refit the kernel after sampling enough new points. The resulting procedure is summarized in \cref{alg:method}.

\begin{algorithm}[!h]
\begin{algorithmic}
\State \textbf{Input:} $P_{\mathrm{obs}}$ - observed histogram
\State \textbf{Options:} $N > 0$ - number of rounds before refitting GP hyperparameters
\State \hphantom{\textbf{Options:}} $m > 0$ - number of pre-training samples
\State \hphantom{\textbf{Options:}} $\eps > 0$ - tolerance
\State \textbf{Output:} $\bx^*$ - parameter estimate
\\
\rule{2.15in}{0.5pt}
\State \textbf{sample} $\bx_1,\ldots,\bx_m \in \mathcal X$
\ForAll{$i = 1,\ldots,m$}
    \State \textbf{compute} $\tilde L(\bx_i)$ by running simulator
    \State $\mathcal D_i \gets \{ \bx_i, \tilde L(\bx_i) \}$
\EndFor
\State \textbf{end for}
\State \textbf{fit} mean and kernel hyperparameters of $\hat L$ \textbf{to} $\mathcal D_{1:m}$
\State $n \gets m$
\Loop
    \State \textbf{maximize} $\alpha(\bx; \hat L \; | \; \mathcal D_{1:n})$
    \State $\bx_{n+1} \gets \arg \max_{\bx} \alpha(\bx; \hat L \; | \; \mathcal D_{1:n})$
    \State \textbf{compute} $\tilde L(\bx_{n+1})$ by running simulator
    \If{$\tilde L(\bx_{n+1}) < \eps$}
        \State $\bx^* \gets \bx_{n+1}$
        \State \textbf{break}
    \EndIf
    \State \textbf{update} $\hat L$ \textbf{with} $\mathcal D_{n+1} = \{ \bx_{n+1}, \tilde L(\bx_{n+1}) \}$
    \State $n \gets n + 1$
    \If{$n-m = 0 \,(\textrm{mod } N)$}
        \State \textbf{refit} kernel hyperparameters of $\hat L$ \textbf{to} $\mathcal D_{1:n}$
    \EndIf
\EndLoop
\State \textbf{end loop} \\
\Return $\bx^*$
\end{algorithmic}
\caption{Bayesian optimization-based parameter estimation}
\label{alg:method}
\end{algorithm}

We often found the value of the loss function $L(\bx)$ to range over orders of magnitude, making it difficult to fit a Gaussian process to $L(\bx)$ in a way that models the function accurately around the minimum. We therefore use a modified version of the loss function given by $L_{\mathrm{log}}(\bx) := \log(1 + L(\bx))$ which satisfies $L_{\mathrm{log}}(\bx) \approx L(\bx)$ for small $L(\bx)$ and grows more slowly for large values of $L$. This significantly reduces the dynamic range of the function modelled by the Gaussian process, improving the efficiency of Bayesian optimization on these problems. 

\subsection{Non-stationary Bayesian optimization}

One issue with the squared exponential kernel commonly used in Gaussian process regression is that it is stationary, that is, the covariance $k(\bx, \bx')$ only depends on the relative difference $\bx - \bx'$. This makes it unsuitable for modelling functions which have different levels of roughness in different parts of parameter space. The loss functions we encountered often displayed a minimum located in a narrow valley surrounded by a large plateau where the loss showed little variation. A Gaussian process with a stationary kernel would either choose very short length scales in order to fit the valley accurately, resulting in a lot of unnecessary uncertainty far away from the minimum and an inefficient optimization procedure due to overexploration, or it would pick large length scales to fit the plateau and treat the observations around the valley as statistical outliers, rendering the optimization routine unable to find the minimum.

Following \cite{martinez-cantin_local_2015} we thus consider a weighted superposition of two independent Gaussian processes, $f = w_g f_g + w_l f_l$ with
\begin{align}
    f_g &\approx \GP(0, k_g(\bx,\bx')) & f_l &\approx \GP(0, k_l(\bx,\bx'))
\end{align}

\noindent and weight functions $w_g(\bx)$, $w_l(\bx)$ to be determined later. This enables us to decompose the Gaussian process into a global component $w_g f_g$ modelling the smooth large-scale behaviour of the loss function and a local component $w_l f_l$ that can fit the function accurately at the minimum. We choose squared exponential kernels $k_g(\bx,\bx')$ and $k_l(\bx,\bx')$ for these two Gaussian process components. The weights are parametrized as
\begin{align}
    w_g(\bx) &= \sqrt{\frac {1}{1 + \nu(\bx)}} & w_l(\bx) &= \sqrt{\frac {\nu(\bx)}{1 + \nu(\bx)}}
\end{align}

\noindent for a nonnegative function $\nu(\bx)$. We set $\nu(\bx)$ to be a squared exponential basis function of the form
\begin{align}
    \nu(\bx) &= \exp\left[- \frac 1 2 (\bx - \bx_\nu)^T \Lambda_\nu (\bx - \bx_\nu) \right]
\end{align}

\noindent for $\Lambda_\nu$ a symmetric positive-definite matrix, chosen to be diagonal in our case, and an anchor point $\bx_\nu$. 

The kernel of the total Gaussian process $f = w_g f_g + w_l f_l$ can be computed to be
\begin{align}
    k(\bx, \bx') = w_g(\bx) w_g(\bx') k_g(\bx, \bx') + w_l(\bx) w_l(\bx') k_l(\bx, \bx')
\end{align}

\noindent As before we fit the hyperparameters by maximum likelihood estimation, constraining $\bx_\nu$ to be the location of the current best observation each time the kernel is refit. This is consistent with our observation that the loss function typically exhibits the largest amount of variation around the minimum.

\section{Results}

\subsection{General setup}
\label{sec:methods_setup}

In all our experiments we chose the $1$-Wasserstein distance with a weighted $\ell^1$ ground metric on $\N^s$. The weight for each species is chosen to be inversely proportional to the mean particle number in the reference distribution, $w_i \propto \E[n_i]^{-1}$; this avoids scenarios where mismatches in abundant species are responsible for the bulk of the Wasserstein loss, leading the optimizer to ignore low-copy number species as it tries to make the largest gains in the loss function.

Our focus on $1$-Wasserstein distances is due to the fact that the $p$-Wasserstein distances are numerically more difficult to compute using the Sinkhorn algorithm for $p > 1$. If the reference distribution can be matched exactly given the right parameters, these parameters will be the global minimum of the Wasserstein loss for all $p \geq 1$, so the choice of $p$ will not affect results. If there is a model mismatch and the reference distribution cannot be reproduced with any set of parameters our algorithm will return parameters minimizing the $p$-Wasserstein distance to that distribution; in this case the minimum will in general depend on the chosen distance. While we have not investigated this issue in detail, we believe that the effect of outliers will be more pronounced for larger $p$ and that $p=1$ is therefore the most stable choice for limited sample sizes.

The steady state distribution of a reaction network can be obtained in two ways:  one can perform simultaneous measurements for a population of cells in the steady state, or one can measure the time average of a single instance of the system over long time scales by the ergodic theorem. Population-level measurements are suitable for experimental data and are commonly realized using e.g. flow cytometry, while in our simulations we prefer to run one instance of the system for long times and compute the time average. Depending on the level of parallelization desired it is possible to run multiple independent simulations for each set of parameters and combine the results, speeding up the inference procedure. All simulations were performed using the Stochastic Simulation Algorithm (SSA) \cite{gillespie_general_1976}.

We check convergence to the steady state distribution by computing Wasserstein distances between the time averages at time points $n T$, $n=1,2,\ldots$, where $T$ is the chosen epoch length. The epoch length $T$ is chosen heuristically such that simulating the system for a few epochs yielded accurate estimates of the steady state distribution for the ground truth system. Simulations are stopped when the distance at two consecutive time points becomes less than 0.02. In order to avoid wasting computation time for parameter settings yielding very bad fits to the observed data we also stop simulations if the distance at two consecutive time points becomes less than $2\%$ of the approximate Wasserstein distance to the observed data. As computing joint Wasserstein distances in multiple dimensions can be expensive, in our simulations we computed the sum of the Wasserstein distances of the marginals when checking for convergence; we found that this did not seem to measurably affect results at the chosen tolerances.

The standard deviation of the observation noise in \cref{eq:obs_noise} for our Gaussian processes is set to $0.03$, which is on the same order of magnitude as the typical measurement error due to finite simulation lengths. In general this hyperparameter can be fit together with the rest of the kernel parameters, but we found this value to work well across experiments. The jitter parameter $\beta$ in \cref{eq:ei} was set to $0.01$, a value commonly used in practice, being e.g.~the default in the Python library \texttt{scikit-learn}.

Since reaction rates are positive and often range over orders of magnitude we use the log reaction rates for inference.

\subsection{Three-stage gene expression model}
\label{sec:gt_3s}

Our first experiment consisted of identifying the parameters in the three-stage model of gene expression found in \cite{shahrezaei_analytical_2008}, described by the following reactions:
\begin{align}
    G &\singlereac{\rho_m} G + M & M &\singlereac{\rho_p} M + P & M &\singlereac{\delta_m} \emptyset \label{eq:3s} \\
    G &\doublereac{\sigma_d}{\sigma_a} G^* & P &\singlereac{1} \emptyset \nonumber
\end{align}

This model consists of four reactant species: a gene in an activated ($G$) and inactivated ($G^*$) form, mRNA ($M$) and protein ($P$). We fix the protein degradation rate to $1$ and perform inference over the remaining five rate constants by observing joint distributions over mRNA and proteins.

\begin{figure}[!h]
    \flushleft
    \hspace{0.35cm}
    \sidesubfloat[]{
    \includegraphics{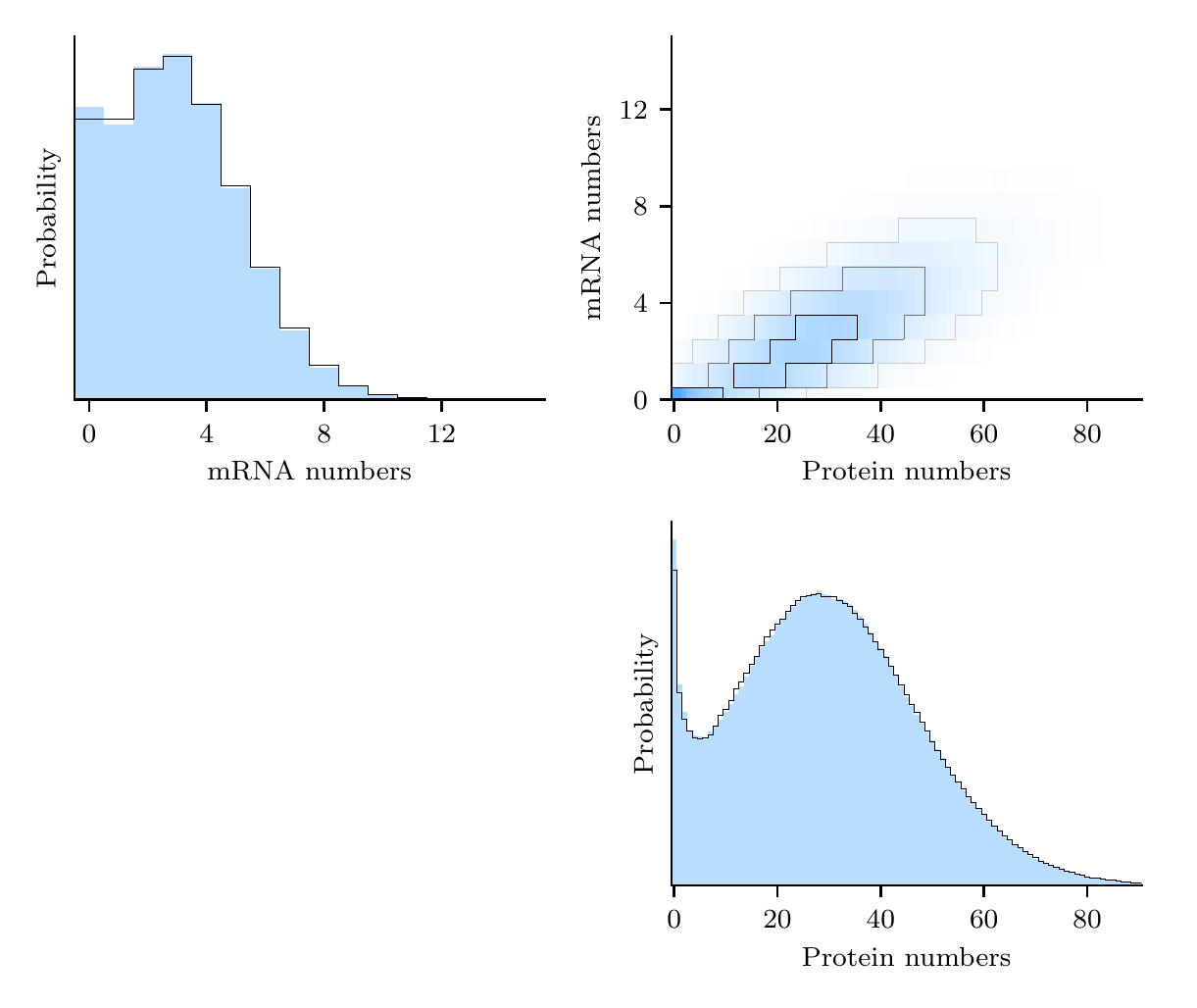}
    \label{fig:GT_3S_a}
    }
    \vspace{-5.15cm}
    
    \hspace{-0.35cm}
    \begin{minipage}{0.585\textwidth}
    \begin{mdframed}[leftline=false,leftmargin=8pt,bottomline=false,innerbottommargin=1.5eM,rightmargin=10pt,innertopmargin=1eM]
    \hspace{-0.45cm}
    \begin{minipage}{0.5\textwidth}
    \sidesubfloat[]{
    \footnotesize
    \setlength{\tabcolsep}{3pt}
    \begin{tabu}{| c @{\hskip 2pt} c  | c @{\hskip 5pt} c @{\hskip 5pt} c @{\hskip 5pt} c @{\hskip 5pt} c |}
        \multicolumn{1}{c}{} \\
        \tabucline{3-7}
        \multicolumn{1}{c}{} & & $\bm{m_{\mathrm{M}}}$ & $\bm{m_{\mathrm{P}}}$ & $\bm{s_{\mathrm{M}}}$ & $\bm{s_{\mathrm{P}}}$ & $\bm{r_{\mathrm{M},\mathrm{P}}}$ \\
        \tabucline{1-7}
        Obs. & $\square$ & 3.00 & 30.1 & 2.16 & 17.6 & 0.73 \\
        BO & $\drawsolidbox{col1}$ & 2.96 & 29.9 & 2.14 & 17.5 & 0.73 \\
        \tabucline{1-7}
    \end{tabu}
    }
    \end{minipage}
    
    \hspace{-0.45cm}
    \vspace{0.25cm}
    \begin{minipage}{0.535\textwidth}
    \hspace{-2.5cm}
    \sidesubfloat[]{
    \footnotesize
    \hspace{-0.5cm}
    \begin{tabu}{| c @{\hskip 2pt} c | c @{\hskip 4pt} c @{\hskip 4pt} c @{\hskip 4pt} c @{\hskip 4pt} c |}
        \multicolumn{1}{c}{} \\
        \tabucline{3-7}
        \multicolumn{1}{c}{} & & $\bm{\rho_m}$ & $\bm{\sigma_d}$ & $\bm{\sigma_a}$ & $\bm{\rho_p}$ & $\bm{\delta_m}$ \\
        \tabucline{1-7}
        \multicolumn{2}{|c|}{Range} & 0.1--10 & 0.01--1 & 0.01--1 & 1--100 & 0.1--10 \\
        \tabucline{1-7}
        GT & $\square$ & 4.00 & 0.20 & 0.60 & 10.0 & 1.00 \\
        BO & $\drawsolidbox{col1}$ & 4.13 & 0.18 & 0.56 & 10.1 & 1.06 \\
        \tabucline{1-7}
    \end{tabu}
    }
    \end{minipage}
    \end{mdframed}
    \end{minipage}
    
    \vspace{0.6cm}
    \caption{Results for the three-stage gene expression model in \cref{eq:3s}. \textbf{(a)}:~Steady steady state distribution over mRNA and protein numbers for the observed data (contours) and the parameters estimated using Bayesian optimization (shaded). \textbf{(b)}: Means~($\bm m$) and standard deviations~($\bm s$) of mRNA and protein numbers as well as their Pearson correlation coefficient~($\bm r$). Both the shape and the moments of the observed distribution are matched by our method. \textbf{(c)}:~Ground truth and estimated parameters for the observed data and the chosen search ranges. The results were obtained after 362 rounds starting with 300 initial samples, where the GP kernel was refit every 75 rounds during optimization.}
    \label{fig:GT_3S}
    
\end{figure}

We fix ground truth values for all parameters (taken from Fig. 3 in \cite{shahrezaei_analytical_2008} with $\gamma = 1$) and use the SSA to obtain a reference steady state distribution. We then apply our method to recover the parameter values based on the observed distribution. The search range for the parameters is set to cover two orders of magnitude per dimension and includes the ground truth values; the results of this experiment can be seen in \cref{fig:GT_3S}. We re-ran the same experiment comparing marginal mRNA and protein numbers (without using the joint distribution) and obtained very similar results, suggesting that it is not always necessary to measure joint distributions to perform parameter inference for the CME if the marginals are fit precisely.
 
A typical run for this experiment lasts about 9--10 hours (Intel Xeon CPU at 3.30 GHz, 32 GB RAM). In comparison, parameter inference for this system using moment equations takes a few minutes. Since the three-stage gene expression model is a linear reaction system it is possible to compute the steady state moments of protein and mRNA numbers exactly without having to resort to moment-closure approximations; parameters can then be estimated exactly using standard optimization methods. While moment-based inference provides a fast and accurate inference method for linear reaction systems, the three-stage gene expression model demonstrates that our method works for nontrivial examples. In the next example, however, we will encounter a nonlinear reaction system for which moment equations perform less well, requiring the use of alternative methods for parameter estimation.

\subsection{Bursty feedback loop}

\label{sec:pfl}

In our second experiment we considered a bursty version of the genetic feedback loop described by the following list of reactions, taken from \cite{grima_steady-state_2012} and \cite{cao_accuracy_2019}:
\begin{align}
    G_u &\singlereac{\rho_u} G_u + k \, P & G_b &\singlereac{\rho_b} G_b + k \, P & & k \sim \mathrm{Geom}(b) \label{eq:pfl} \\
    G_u + P &\doublereac{\sigma_b}{\sigma_u} G_b & P &\singlereac{1} \emptyset & & \nonumber
\end{align}

\noindent This system describes a protein which can bind to its gene and hence influence its own transcription rate. The number of proteins produced at each translation event follows a geometric distribution with mean $b$,
\begin{align*}
    p(k) = \frac{b^k}{(1 + b)^{k+1}} \qquad (k \geq 0)
\end{align*}

\noindent which is a common approximation of mRNA-mediated protein production when the lifetime of mRNA is very short compared to the mean protein lifetime \cite{friedman_linking_2006}.

The system \labelcref{eq:pfl} displays different types of behaviour depending on whether $\rho_u > \rho_b$ (negative feedback) or $\rho_u < \rho_b$ (positive feedback). In \cite{cao_accuracy_2019} the authors compare different moment closure approximations as well as the Linear Mapping Approximation \cite{cao_linear_2018} and show that for the negative feedback loop it is possible to efficiently obtain accurate parameter estimates using a suitable moment closure scheme. In this section we will focus on the positive feedback case, which we found to be more challenging for the approach presented in \cite{cao_accuracy_2019}.

\begin{figure}[!tb]
    \centering
    \includegraphics{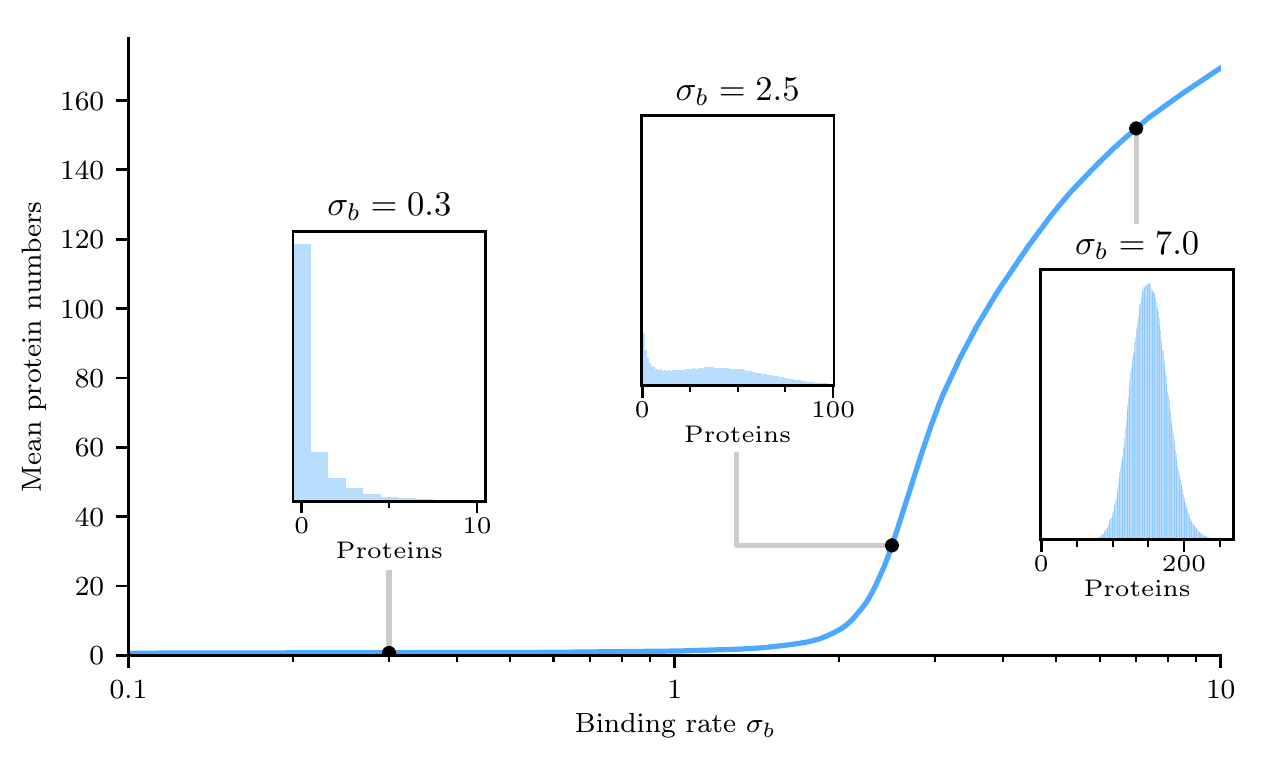}
    \caption{Mean protein numbers plotted against the binding rate $\sigma_b$ for the bursty positive feedback loop. There is a sharp increase around $\sigma_b \approx 2$ as the system switches from being in the inactivated state most of the time to the activated state. The steady state distributions differ qualitatively depending on $\sigma_b$, with the intermediate region characterized by bimodal protein number distributions. The values of the remaining parameters are $\rho_u=0.3$, $\sigma_u=400$, $\rho_b=105$, $b=2$.}
    \label{fig:PFL_means}
\end{figure}

Positive feedback in this system can result in strong sensitivity to parameter values (see \cref{fig:PFL_means}). We chose parameters that resulted in the gene spending non-negligible amounts of time in both the bound and the unbound state. This regime (which we shall call the intermediate regime) is characterized by a bimodal steady state distribution over protein numbers which changes rapidly with $\sigma_b$ and $\rho_b$.

In order to test how well moment closure methods can approximate the positive feedback loop we applied six different moment closure schemes from \cite{cao_accuracy_2019}: conditional derivative matching and conditional Gaussian \cite{soltani_conditional_2015}, both conditioned on either the bound or the unbound states of the gene (denoted CDM1 and CDM2, resp. CG1 and CG2), as well as unconditional Gaussian (Gauss) and derivative matching (DM) \cite{singh_derivative_2007}. We also considered the Linear Mapping Approximation (LMA) \cite{cao_linear_2018} as it yields a set of moment equations that can be solved directly, similar to classical moment closure schemes. In addition we tested the conditional negative binomial approximation, again conditioned on both the bound and the unbound state of the gene, which we respectively denote CNB1 and CNB2.

\begin{figure}[!tb]
\floatbox[{\capbeside\thisfloatsetup{capbesideposition={right,center},capbesidewidth=0.325\textwidth}}]{figure}[\FBwidth]
{\caption{Eight different moment closure schemes and the LMA applied to the positive feedback loop, using ground truth the parameters given in \cref{fig:PFL_bursty}. CG1 and DM failed to yield a solution predicting a positive mean. Even the best approximation (CNB1) is more than $30\%$ off in its estimate of both the mean and the standard deviation of protein numbers.}\label{fig:PFL_mean_preds}}
{
    \includegraphics{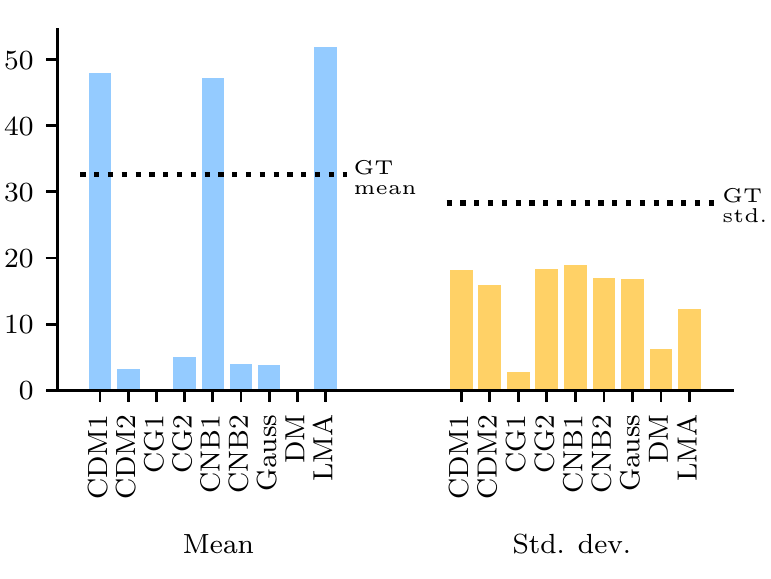}
}
\end{figure}

We found that none of the nine methods tested were able to accurately predict mean and standard deviation of the protein number distribution for our system (\cref{fig:PFL_mean_preds}). While many were able to model the system outside the intermediate regime, the presence of large fluctuations in that regime significantly decreased the accuracy of the methods. The closed moment equations are nonlinear and usually admit multiple solutions, yielding complex, negative or outlandishly high predictions for moments. This is similar to the scenario tested in \cite{schnoerr_comparison_2015} where the considered moment closure schemes failed to yield unique solutions in general. We found that with the exception of the LMA all tested moment closure methods broke down in different parts of the intermediate regime, and while the LMA itself always yields an interpretable solution it returns inaccurate predictions of moments for our system (cf.~\cref{fig:PFL_mean_preds}).

Given that none of the moment closure methods used for the negative feedback case capture the intermediate regime in the positive feedback case one has to rely on alternative inference methods for our problem. The complicated dependence of the steady state on the parameters makes this task challenging in general. In the intermediate regime very small changes in $\sigma_b$ or $\rho_b$ will typically lead to large changes in the steady state distribution, while in the regime where the gene is mostly unbound the system will virtually be independent of $\sigma_b$ (cf. \cref{fig:PFL_means}). Hence the loss landscape looks very different at different points in parameter space, causing problems for both global and local optimization approaches. A grid search for $\sigma_b$ for example would need to sample values at very short intervals in order to find the correct value in the intermediate regime, while a local optimization routine would likely get stuck if initialized in a region where the loss function is flat. 

We tested our method on this problem by performing joint inference over $\sigma_b$, $\rho_u$ and $\sigma_u$ in the intermediate regime, based on observing the (marginal) protein number distribution. The results are shown in \cref{fig:PFL_bursty}. We recovered a set of parameters that yield a steady state protein distribution closely matching the input data and which themselves are in broad agreement with the ground truth. We suspect that the observed discrepancy of about $10-15\%$ is due to local non-identifiability of the parameters, consistent with previous observations for the negative feedback loop \cite{cao_accuracy_2019}. We remark that changing only $\sigma_b$ from its ground truth value does not yield a similar steady state distribution as can be seen in \cref{fig:PFL_means}, suggesting that the non-identifiability involves trade-offs between different parameters.

The time required for a typical run of our algorithm with this reaction system was 3.5 hours, which is noticeably faster than the three-stage gene expression model due to the reduced number of simulations needed for convergence.

\setcounter{figure}{4}
\begin{figure}[!tb]
    \floatbox[{\capbeside\thisfloatsetup{capbesideposition={right,center},capbesidewidth=0.45\textwidth}}]{figure}[\FBwidth]
    {
    \caption{Results for the bursty positive feedback loop. \textbf{(a)}:~Steady state distribution over protein numbers for the observed data (contours) and the estimated parameters (shaded). \textbf{(b)}: Means and standard deviations of the two distributions. \textbf{(c)}:~Ground truth and estimated parameters for the observed data. The results were obtained after 130 rounds starting with 75 initial samples, where the kernel was refit every 25 rounds during optimization. The remaining parameters are $\rho_b = 105$, $b = 2$ (cf.~\cref{fig:PFL_means}). \label{fig:PFL_bursty}}}
    {
    \hspace{-0.05\textwidth}
    \begin{minipage}{0.5\textwidth}
    
    \flushleft
    \begin{minipage}[t]{0.5\textwidth}
    \sidesubfloat[]{
        \hspace{-0.5cm}
        \includegraphics{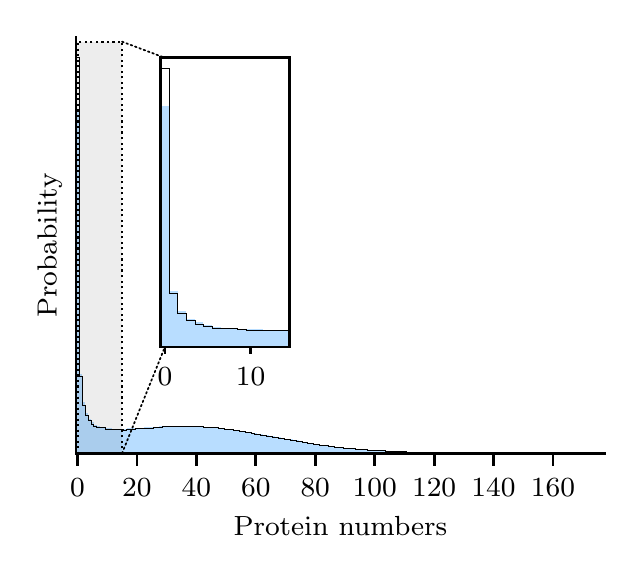}
    }
    \end{minipage}
    ~
    \begin{minipage}[t]{0.4\textwidth}
    \footnotesize
    \vspace{0.3cm}
    \setlength{\tabcolsep}{3pt}
    \sidesubfloat[]{
    \hspace{-0.8cm}
    \begin{tabu}{| c @{\hskip 2pt} c | c @{\hskip 5pt} c  |}
        \multicolumn{1}{c}{} \\
        \tabucline{3-4}
        \multicolumn{1}{c}{} & & $\bm{m_{\mathrm{P}}}$ & $\bm{s_{\mathrm{P}}}$ \\
        \tabucline{1-4}
        Obs. & $\square$ & 32.6 & 28.0 \\
        BO & $\drawsolidbox{col1}$ & 33.1 & 28.3 \\
        \tabucline{1-4}
    \end{tabu}
    }
    \end{minipage}

    \begin{minipage}{0.5\textwidth}
    \centering
    \sidesubfloat[]{

    \footnotesize
    \hspace{0.35cm}
    \vspace{-1cm}
    \begin{tabu}{| c @{\hskip 2pt} c | c @{\hskip 4pt} c @{\hskip 4pt} c |}
        \multicolumn{1}{c}{} \\
        \tabucline{3-5}
        \multicolumn{1}{c}{} & & $\bm{\rho_u}$ & $\bm{\sigma_u}$ & $\bm{\sigma_b}$ \\
        \tabucline{1-5}
        \multicolumn{2}{|c|}{Range} & 0.15--15 & 10-1000 & 0.1--10  \\
        \tabucline{1-5}
        GT & $\square$ & 0.30 & 400 & 2.50 \\
        BO & $\drawsolidbox{col1}$ & 0.34 & 355 & 2.22  \\
        \tabucline{1-5}
    \end{tabu}
    }
    \end{minipage}
    \end{minipage}
    }
\end{figure}

\FloatBarrier 
\subsection{Further experiments}

\label{sec:exp}

In order to gain a better understanding of the relationship between parameter estimation using Wasserstein distances and existing methods we performed additional tests, investigating the accuracy of our method for limited sample sizes and for variations of the bursty feedback loop from \cref{sec:pfl}.

In practical applications one does not usually have direct access to the steady-state distribution of a reaction system due to the finite number of cells that can be measured at once. The reference distribution will therefore always be an empirical measure of the ground truth, where sample sizes of at least $100$ are common for population snapshot data. It is therefore important to understand how well our method performs when the input distribution is such an empirical distribution.  

We tested our method on the three-stage gene expression model, the positive feedback loop and negative feedback loop considered in \cite{cao_accuracy_2019}, taking as inputs empirical estimates of the ground truth distribution for different sample sizes. Results are displayed in \cref{tbl:comparison_emp_fl,tbl:comparison_emp_gt}. In all cases the estimated parameters yield steady-state distributions that are similar to both the empirical (input) distribution and the original ground truth data, where the quality of the fit improves with the sample size. We found that larger sample sizes seem to be necessary where multiple species are observed simultaneously. The estimated steady-state distributions try to match the means and standard variations of the empirical data, which is not surprising as convergence in any of the Wasserstein distance generally implies convergence of the moments \cite{villani_optimal_2009}. This suggests that Wasserstein-based inference behaves similarly to moment-based inference in the context of limited sample sizes.

\begin{table}[tb]
\footnotesize
\begin{tabu}{| c | c | c | c | c | c | c | c | c | c |}
    \tabucline{1-10}
    \multicolumn{10}{|c|}{$\bm{N=500}$} \\
    \tabucline{1-10}
    \multicolumn{5}{|c|}{\textbf{Obs.}} & \multicolumn{5}{c|}{\textbf{BO}} \\
    \tabucline{1-10}
    $\bm{m_M}$ & $\bm{m_P} $ & $\bm{s_M}$ & $\bm{s_P} $ & $\bm{r_{M,P}}$ & $\bm{m_M}$ & $\bm{m_P} $ & $\bm{s_M}$ & $\bm{s_P} $ & $\bm{r_{M,P}}$ \\
    \tabucline{1-10}
    2.89 & 29.9 & 2.05 & 17.2 & 0.69 & 2.92 & 30.3 & 2.07 & 16.2 & 0.67 \\
    \tabucline{1-10}
    \multicolumn{10}{c}{} \\
    \tabucline{1-10}
    \multicolumn{10}{|c|}{$\bm{N=1000}$} \\
    \tabucline{1-10}
    \multicolumn{5}{|c|}{\textbf{Obs.}} & \multicolumn{5}{c|}{\textbf{BO}} \\
    \tabucline{1-10}
    $\bm{m_M}$ & $\bm{m_P} $ & $\bm{s_M}$ & $\bm{s_P} $ & $\bm{r_{M,P}}$ & $\bm{m_M}$ & $\bm{m_P} $ & $\bm{s_M}$ & $\bm{s_P} $ & $\bm{r_{M,P}}$ \\
    \tabucline{1-10}
    2.98 & 29.7 & 2.16 & 17.4 & 0.71 & 2.99 & 29.7 & 2.03 & 17.2 & 0.76\\
    \tabucline{1-10}
\end{tabu}
\caption{Marginal means and standard deviations over mRNA and protein counts as well as their correlation coefficients for the three-stage gene expression network given empirical input distributions with sample sizes $N=500$ and $1000$. Our method was run for 325 rounds starting with 300 initial samples, with the remaining setup as in \cref{sec:gt_3s}.}

\label{tbl:comparison_emp_gt}
\end{table}

{\centering
\begin{table}[tb]
\footnotesize
\begin{tabu}{|@{\hspace{2pt}} c @{\hspace{2pt}}| c | c | c | c | c | c | c | c | c | c | c | c |}
    \tabucline{1-13}
    \multirow{3}{*}{\textbf{System}} & \multicolumn{4}{c|}{$\bm{N = 100}$} & \multicolumn{4}{c|}{$\bm{N = 500}$} & \multicolumn{4}{c|}{$\bm{N = 1000}$} \\
    \tabucline{2-13}
    & \multicolumn{2}{c|}{\textbf{Obs.}} & \multicolumn{2}{c|}{\textbf{BO}} & \multicolumn{2}{c|}{\textbf{Obs.}} & \multicolumn{2}{c|}{\textbf{BO}} & \multicolumn{2}{c|}{\textbf{Obs.}} & \multicolumn{2}{c|}{\textbf{BO}} \\
    \tabucline{2-13}
    & $\bm{m_P}$ & $\bm{s_P} $ & $\bm{m_P}$ & $\bm{s_P} $ & $\bm{m_P}$ & $\bm{s_P} $  & $\bm{m_P}$ & $\bm{s_P} $ & $\bm{m_P}$ & $\bm{s_P} $ & $\bm{m_P}$ & $\bm{s_P} $ \\
    \tabucline{1-13}
    PFL & 30.1 & 28.6 & 30.6 & 28.2 & 30.0 & 27.5 & 29.1 & 26.7 & 33.2 & 27.7 & 32.4 & 28.4 \\
    NFL & 25.2 & 22.0 & 25.0 & 21.0 & 28.6 & 19.1 & 28.5 & 19.1 & 28.0 & 19.7 & 28.1 & 19.2 \\
    \tabucline{1-13}
\end{tabu}
\caption{Mean and standard deviation over protein counts for the three-stage gene expression network given empirical input distributions with sample sizes $N=100$, $500$ and $1000$. Our method was run for 150 rounds starting with 75 initial samples, with the remaining setup as in \cref{sec:pfl}.}

\label{tbl:comparison_emp_fl}
\end{table}
}

Existing inference methods for the CME for steady-state data fall broadly into the category of moment-based inference methods and direct, likelihood-based methods. The direct approach in \cite{neuert_systematic_2013,munsky_distribution_2018} differs from our proposal in that it uses the Finite State Projection algorithm (FSP) to compute the steady-state distribution. The computational time required by the FSP depends on the number of states considered, which in turn is determined by the desired accuracy and typically increases exponentially in the number of species. As the time complexity of the FSP is cubic in the number of states it does not scale as well as the simulation-based approach for nontrivial reaction systems and we therefore concentrated on moment-based inference in the following comparison.


To investigate the accuracy of moment-based inference for a nonlinear reaction network we evaluated the performance of both approaches on four variations of the feedback loop considered in \cref{sec:pfl} with different parameters. The first is the negative feedback loop (NFL) considered in \cite{cao_accuracy_2019}, and the remainder was obtained by randomly drawing parameters for the entire reaction system and taking the first three parameter settings that yielded nontrivial steady-state distributions ($m_P > 5$), which we call R1, R2 and R3. The chosen parameters and search ranges are given in \cref{tbl:comparison_data}. For moment-based inference we tested all of the 9 moment-closure schemes listed in \cref{sec:pfl}. For our method the setup is the same as in \cref{sec:pfl}, each run consisting of 150 optimization rounds starting with 75 initial samples.

We found moment-based inference to work well for the negative feedback loop, with all but two of the listed moment-closure schemes achieving accuracy comparable to our results (see \cref{sec:apdx_results}). In some of these cases the mean and variance of the steady-state distribution were matched accurately, but the shape of the distributions differed perceptibly, which suggests that means and variances are not always enough to uniquely characterise steady-state distributions (see \cref{fig:mom_disc}). Results for the remaining three networks were mixed: while some moment-closure schemes worked well for either R2 or R3, many introduced a perceptible bias into the results, and for R1 the resulting standard deviation was consistently too low. We report the results of our experiments in \cref{tbl:comparison_mom,tbl:comparison_param}, choosing the most accurate moment-closure scheme for each system as a representative of moment-based inference. Results for the remaining moment-closure schemes are given in Appendix \labelcref{sec:apdx_results}.

\setcounter{figure}{7}
\begin{figure}[htb]
    \floatbox[{\capbeside\thisfloatsetup{capbesideposition={right,center},capbesidewidth=0.45\textwidth}}]{figure}[\FBwidth]
    {
    \caption{Moments do not always distinguish steady-state distributions in practice. Shown are steady-state distributions for the original negative feedback loop (contours) and for a modification (shaded). While the two distributions are perceptibly different their means and standard deviations differ by less than $1\%$. The modified parameters are $\rho_u = 14.5$, $\sigma_u = 0.33$, $\sigma_b = 0.0048$.} 
    \label{fig:mom_disc}}
    {
    \hspace{-0.05\textwidth}
    \begin{minipage}{0.5\textwidth}
    
    \flushleft
    \begin{minipage}[t]{0.5\textwidth}
        \hspace{-0.5cm}
        \includegraphics{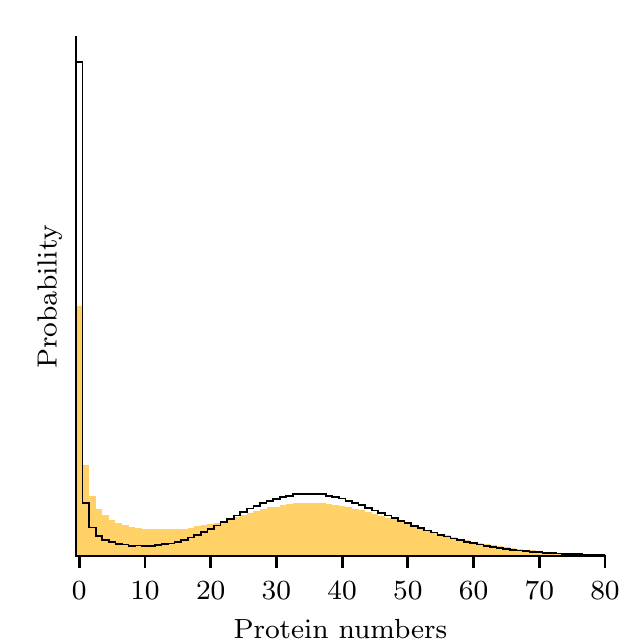}
    \end{minipage}
    ~
    \begin{minipage}[t]{0.4\textwidth}
    \footnotesize
    \vspace{0.5cm}
    \setlength{\tabcolsep}{3pt}
    \hspace{-0.5cm}
    \begin{tabu}{| c @{\hskip 2pt} c | c @{\hskip 5pt} c  |}
        \multicolumn{1}{c}{} \\
        \tabucline{3-4}
        \multicolumn{1}{c}{} & & $\bm{m_{\mathrm{P}}}$ & $\bm{s_{\mathrm{P}}}$ \\
        \tabucline{1-4}
        NFL & $\square$ & 28.1 & 19.5 \\
        Mod. & $\drawsolidboxtwo{col2}$ & 28.2 & 19.5 \\
        \tabucline{1-4}
    \end{tabu}
    \end{minipage}

    \end{minipage}
    \vspace{-0.5cm}
    }
\end{figure}

Parameter values returned by either method frequently differed from their ground truth values even when the steady-state distributions matched up. This suggests that parameter identifiability is a common problem for this class of reaction networks (cf.~\cite{cao_accuracy_2019}). 

We point out that the accuracy of our method matches that of moment-based inference in all cases despite the fact that our approach is model-agnostic and can be applied to any reaction system without modifications, whereas all but two of the moment-closure schemes we used are tailored to the genetic feedback loop. The wealth of moment-closure schemes available for this reaction system is due to its structure as a genetic switch and conditional moment-closure schemes or the LMA \cite{cao_linear_2018} do not in general have equivalents for other types of reaction networks. 

Runtimes for our method were on the order of one hour, except for the negative feedback loop (2.5 hours) and the positive feedback loop (3.5 hours). Moment-based inference was very fast, usually taking less than a minute per (convergent) run. We found that some moment-closure methods ran into numerical issues during optimization for some of the systems considered, consistent with the issues we reported for the positive feedback loop in \cref{sec:pfl} and leading to drastically increased runtimes.

\begin{table}[H]
\footnotesize

    \begin{tabu}{| c | c | r @{--} l | c | r @{--} l | c | r @{--} l | c | c |}
        \tabucline{1-12}
        \textbf{System} & $\bm{\rho_u}$ & \multicolumn{2}{c|}{\textbf{Range}} & $\bm{\sigma_u}$ & \multicolumn{2}{c|}{\textbf{Range}} & $\bm{\sigma_b}$ &\multicolumn{2}{c|}{\textbf{Range}} & $\bm{\rho_b}$ & $\bm{b}$ \\
        \tabucline{1-12}
        NFL & 13 & 1 & 100 & 0.1 & 0.01 & 1 & 0.001 & $10^{-4}$ & 0.01 & 0 & 3 \\
        R1 & 1.3 & 1 & 100 & 42 & 1 & 100 & 1.5 & 0.1 & 10 & 63 & 1.5 \\
        R2 & 9.3 & 0.1 & 10 & 1 & 0.1 & 10 & 0.7 & 0.1 & 10 & 0.4 & 8.3 \\
        R3 & 1 & 0.1 & 10 & 355 & 10 & 1000 & 7.9 & 0.1 & 10 & 20.1 & 2.4 \\
        \tabucline{1-12}
    \end{tabu}
    
\caption{Ground truth parameter values and search ranges for the additional reaction networks tested in this section.}

\label{tbl:comparison_data}
\end{table}

\begin{table}[H]
\footnotesize
\begin{tabu}{| c | c | c | c | c | c | c | c |}
    \tabucline{1-8}
    \multirow{2}{*}{\textbf{System}} & \multicolumn{2}{c|}{\textbf{GT}}  & \multicolumn{2}{c|}{\textbf{BO}} & \multicolumn{3}{c|}{\textbf{MBI}} \\
    \tabucline{2-8}
    & $\bm{m_P}$ & $\bm{s_P} $ & $\bm{m_P}$ & $\bm{s_P} $ & $\bm{m_P}$ & $\bm{s_P} $ & \textbf{Method} \\
    \tabucline{1-8}
    PFL & 32.6 & 28.0 & 33.1 & 28.3 & 31.9 & 21.4 & CG1 \\ 
    NFL & 27.9 & 19.5 & 27.9 & 19.5 & 28.5 & 19.3 & CG2 \\ 
    R1 & 66.4 & 16.1 & 66.4 & 15.8 & 66.5 & 13.5 & CG2 \\ 
    R2 & 11.4 & 10.1 & 11.3 & 9.9 & 11.6 & 10.1 & CG2 \\ 
    R3 & 9.9 & 9.3 & 9.7 & 9.2 & 10.7 & 9.0 & CG1 \\ 
    \tabucline{1-8}
\end{tabu}
\caption{Mean and standard deviation over protein counts for the ground truth (GT), our method (BO) and moment-based inference (MBI), as well as the corresponding moment-closure scheme in the last column.}

\label{tbl:comparison_mom}
\end{table}

\begin{table}[H]
\footnotesize
\begin{tabu}{| c | c | c | c | c | c | c | c | c | c |}
    \tabucline{1-10}
    \multirow{2}{*}{\textbf{System}} & \multicolumn{3}{c|}{\textbf{GT}}  & \multicolumn{3}{c|}{\textbf{BO}} & \multicolumn{3}{c|}{\textbf{MBI}} \\
    \tabucline{2-10}
    & $\bm{\rho_u}$ & $ \bm{\sigma_u} $ & $\bm{\sigma_b}$ & $\bm{\rho_u}$ & $ \bm{\sigma_u} $ & $\bm{\sigma_b}$ & $\bm{\rho_u}$ & $ \bm{\sigma_u} $ & $\bm{\sigma_b}$ \\
    \tabucline{1-10}
    PFL & 0.3 & 400 & 2.5 & 0.34 & 355 & 2.22 & 1.41 & 1000.0 & 5.45 \\
    NFL & 13 & 0.1 & 0.001 & 12.7 & 0.055 & 0.0006 & 12.9 & 0.071 & 0.0007 \\ 
    R1 & 1.3 & 42 & 1.5 & 22.2 & 15.1 & 0.28 & 40.2 & 50.2 & 0.17 \\ 
    R2 & 9.3 & 1 & 0.7 & 1.54 & 6.43 & 0.10 & 1.49 & 10.0 & 0.10 \\ 
    R3 & 1 & 355 & 7.9 & 1.14 & 88.2 & 1.87 & 1.36 & 326.4 & 6.67  \\ 
    \tabucline{1-10}
\end{tabu}
\caption{Parameters estimated by our method and using moment-based inference (see~\cref{tbl:comparison_mom}).}

\label{tbl:comparison_param}
\end{table}

\FloatBarrier 

\section{Conclusion}

We presented a general-purpose method for parameter estimation for simulator-based stochastic reaction networks. We computed Wasserstein distances between steady state distributions to quantify the discrepancy between observed data and simulator output for different parameters and constructed a probabilistic model of the Wasserstein distance at unexplored parameter settings by training a Gaussian process with these data. Bayesian optimization was applied in order to (i) iteratively choose parameter settings that are likely to be close to the optimum, (ii) evaluate the system at the chosen parameters and (iii) update the model until the results are consistent with observations. We applied our method to two inverse problems for the Chemical Master Equation: a five-dimensional problem based on the classical three-stage gene expression model, and a three-dimensional problem based on a genetic feedback loop, recovering data consistent with observations in both cases. We performed further experiments that demonstrated that our method compares favourably with moment-based inference in terms of accuracy.

Our method only relies on having access to a simulator and does not require the computation of likelihoods. It is thus especially suited to models such as Brownian Dynamics which can be sampled from but which are in general not tractable analytically. Given the fact that many simulator models are expensive to evaluate, Bayesian optimization provides an effective method to perform inference with a limited number of simulations. To our knowledge there is no previous literature on efficient parameter estimation for Brownian Dynamics or related models, and we hope that the approach presented will provide a first stepping stone in this direction.

Bayesian optimization has previously been used for likelihood-free inference e.g.~in  \cite{gutmann_bayesian_2016}. One potential limitation of global optimization approaches like Bayesian optimization is that they are often difficult to apply to high-dimensional problems. The number of evaluations needed until convergence usually scales with the dimension of the parameter space, reducing their usability for problems with many parameters. The effectiveness of Bayesian optimization in particular depends strongly on the ability to model complex high-dimensional functions using a Gaussian process given a limited amount of evaluations, a nontrivial task. We are positive that continuing research on non-stationary Bayesian optimization methods \cite{martinez-cantin_local_2015,snoek_input_2014} will enable us to deal with these problems more effectively in the future.

In this paper we used Wasserstein distances to match distributions, one problem with these being that the computations can become expensive for large histogram dimensions. There are multiple possible remedies for this. The Sinkhorn algorithm consists mostly of repeated matrix-vector operations and can be easily implemented on GPUs, potentially resulting in significant speed gains compared to using a CPU. For systems with large particle numbers one can coarse-grain the histogram by binning particle numbers and approximating the Wasserstein distances using the coarsened histograms. Finally one can compute Wasserstein distances between lower-dimensional marginals separately and minimize the sum of the distances. While this method may potentially lose information about correlations between different species we have observed it to yield near identical results for the three-stage gene expression model. Since it is difficult to experimentally measure joint distributions of several distinct species, fitting marginalized distributions is a reasonable approach in these situations and we found it to perform well in practice.

\section*{Data availability}

The code for this project is available at \url{https://github.com/kaandocal/wasserstein_inference}.

\section*{Acknowledgments}

We would like to thank our reviewers for their suggestions and careful reading of the manuscript. This work was supported in part by the EPSRC Centre for Doctoral Training in Data Science, funded by the UK Engineering and Physical Sciences Research Council (grant EP/L016427/1) and the University of Edinburgh.

\section*{References}

\printbibliography[heading=none]

@article{golightly_bayesian_2006,
	title = {Bayesian Sequential Inference for Stochastic Kinetic Biochemical Network Models},
	volume = {13},
	issn = {1066-5277, 1557-8666},
	url = {http://www.liebertpub.com/doi/10.1089/cmb.2006.13.838},
	doi = {10.1089/cmb.2006.13.838},
	pages = {838--851},
	number = {3},
	journaltitle = {Journal of Computational Biology},
	shortjournal = {J. Comput. Bio.},
	author = {Golightly, Andrew and Wilkinson, Darren J.},
	urldate = {2018-09-21},
	date = {2006-04},
	langid = {english}
}

@article{golightly_bayesian_2011,
	title = {Bayesian parameter inference for stochastic biochemical network models using particle {Markov} chain {Monte Carlo}},
	volume = {1},
	issn = {2042-8898},
	url = {https://www.ncbi.nlm.nih.gov/pmc/articles/PMC3262293/},
	doi = {10.1098/rsfs.2011.0047},
	pages = {807--820},
	number = {6},
	journaltitle = {Interface Focus},
	shortjournal = {Interface Focus},
	author = {Golightly, Andrew and Wilkinson, Darren J.},
	urldate = {2018-09-21},
	date = {2011-12-06},
	pmid = {23226583},
	pmcid = {PMC3262293}
}

@article{shahrezaei_analytical_2008,
	title = {Analytical distributions for stochastic gene expression},
	volume = {105},
	rights = {© 2008 by The National Academy of Sciences of the {USA}},
	issn = {0027-8424, 1091-6490},
	url = {http://www.pnas.org/content/105/45/17256},
	doi = {10.1073/pnas.0803850105},
	pages = {17256--17261},
	number = {45},
	journaltitle = {Proceedings of the National Academy of Sciences},
	shortjournal = {{PNAS}},
	author = {Shahrezaei, Vahid and Swain, Peter S.},
	urldate = {2018-11-12},
	date = {2008-11-11},
	langid = {english},
	pmid = {18988743}
}

@article{gillespie_general_1976,
	title = {A general method for numerically simulating the stochastic time evolution of coupled chemical reactions},
	volume = {22},
	issn = {00219991},
	url = {http://linkinghub.elsevier.com/retrieve/pii/0021999176900413},
	doi = {10.1016/0021-9991(76)90041-3},
	pages = {403--434},
	number = {4},
	journaltitle = {Journal of Computational Physics},
	shortjournal = {J. Comput. Phys.},
	author = {Gillespie, Daniel T},
	urldate = {2018-11-28},
	date = {1976-12},
	langid = {english}
}

@article{martinez-cantin_local_2015,
	title = {Local nonstationarity for efficient {Bayesian} optimization},
	url = {http://arxiv.org/abs/1506.02080},
	author = {Martinez-Cantin, Ruben},
	urldate = {2019-04-05},
	date = {2015-06-05},
	eprinttype = {arxiv},
	eprint = {1506.02080}
}

@article{munsky_finite_2006,
	title = {The {Finite State Projection} algorithm for the solution of the {Chemical Master Equation}},
	volume = {124},
	issn = {0021-9606},
	url = {https://aip.scitation.org/doi/full/10.1063/1.2145882},
	doi = {10.1063/1.2145882},
	pages = {044104},
	number = {4},
	journaltitle = {The Journal of Chemical Physics},
	shortjournal = {J. Chem. Phys.},
	author = {Munsky, Brian and Khammash, Mustafa},
	urldate = {2019-02-13},
	date = {2006-01-25}
}

@article{cao_accuracy_2019,
	title = {Accuracy of parameter estimation for auto-regulatory transcriptional feedback loops from noisy data},
	volume = {16},
	issn = {1742-5689, 1742-5662},
	url = {http://www.royalsocietypublishing.org/doi/10.1098/rsif.2018.0967},
	doi = {10.1098/rsif.2018.0967},
	pages = {20180967},
	number = {153},
	journaltitle = {Journal of the Royal Society Interface},
	shortjournal = {J. R. Soc. Interface},
	author = {Cao, Zhixing and Grima, Ramon},
	urldate = {2019-06-15},
	date = {2019-04-26},
	langid = {english}
}

@article{schwanhausser_global_2011,
	title = {Global quantification of mammalian gene expression control},
	volume = {473},
	rights = {2011 Nature Publishing Group},
	issn = {1476-4687},
	url = {https://www.nature.com/articles/nature10098},
	doi = {10.1038/nature10098},
	pages = {337--342},
	journaltitle = {Nature},
	author = {Schwanhäusser, Björn and Busse, Dorothea and Li, Na and Dittmar, Gunnar and Schuchhardt, Johannes and Wolf, Jana and Chen, Wei and Selbach, Matthias},
	urldate = {2019-06-15},
	date = {2011-05},
	langid = {english}
}

@book{van_kampen_stochastic_2007,
	edition = {3},
	title = {Stochastic Processes in Physics and Chemistry},
	isbn = {978-0-444-52965-7},
	url = {https://linkinghub.elsevier.com/retrieve/pii/B9780444529657X50004},
	publisher = {Elsevier},
	author = {van Kampen, N.G.},
	urldate = {2019-06-15},
	date = {2007},
	langid = {english},
	doi = {10.1016/B978-0-444-52965-7.X5000-4}
}

@article{munsky_distribution_2018,
	title = {Distribution shapes govern the discovery of predictive models for gene regulation},
	volume = {115},
	rights = {Copyright © 2018 the Author(s). Published by {PNAS}.. https://creativecommons.org/licenses/by-nc-nd/4.0/This open access article is distributed under Creative Commons Attribution-{NonCommercial}-{NoDerivatives} License 4.0 ({CC} {BY}-{NC}-{ND}).},
	issn = {0027-8424, 1091-6490},
	url = {https://www.pnas.org/content/115/29/7533},
	doi = {10.1073/pnas.1804060115},
	pages = {7533--7538},
	number = {29},
	journaltitle = {Proceedings of the National Academy of Sciences},
	shortjournal = {{PNAS}},
	author = {Munsky, Brian and Li, Guoliang and Fox, Zachary R. and Shepherd, Douglas P. and Neuert, Gregor},
	urldate = {2019-06-06},
	date = {2018-07},
	langid = {english},
	pmid = {29959206}
}

@article{ruess_moment-based_2015,
	title = {Moment-based methods for parameter inference and experiment design for stochastic biochemical reaction networks},
	volume = {25},
	issn = {1049-3301},
	url = {http://doi.acm.org/10.1145/2688906},
	doi = {10.1145/2688906},
	pages = {8:1--8:25},
	number = {2},
	journaltitle = {{ACM} Trans. Model. Comput. Simul.},
	author = {Ruess, Jakob and Lygeros, John},
	urldate = {2019-06-06},
	date = {2015-02}
}

@article{neuert_systematic_2013,
	title = {Systematic identification of signal-activated stochastic gene regulation},
	volume = {339},
	rights = {Copyright © 2013, American Association for the Advancement of Science},
	issn = {0036-8075, 1095-9203},
	url = {https://science.sciencemag.org/content/339/6119/584},
	doi = {10.1126/science.1231456},
	pages = {584--587},
	number = {6119},
	journaltitle = {Science},
	author = {Neuert, Gregor and Munsky, Brian and Tan, Rui Zhen and Teytelman, Leonid and Khammash, Mustafa and Oudenaarden, Alexander van},
	urldate = {2019-06-06},
	date = {2013-02},
	langid = {english},
	pmid = {23372015}
}

@article{cinquemani_identifiability_2018,
	title = {Identifiability and reconstruction of biochemical reaction networks from population snapshot data},
	volume = {6},
	rights = {http://creativecommons.org/licenses/by/3.0/},
	url = {https://www.mdpi.com/2227-9717/6/9/136},
	doi = {10.3390/pr6090136},
	pages = {136},
	number = {9},
	journaltitle = {Processes},
	author = {Cinquemani, Eugenio},
	urldate = {2019-06-06},
	date = {2018-09},
	langid = {english}
}

@article{schnoerr_comparison_2015,
	title = {Comparison of different moment-closure approximations for stochastic chemical kinetics},
	volume = {143},
	issn = {0021-9606, 1089-7690},
	url = {http://aip.scitation.org/doi/10.1063/1.4934990},
	doi = {10.1063/1.4934990},
	pages = {185101},
	number = {18},
	journaltitle = {The Journal of Chemical Physics},
	shortjournal = {J. Chem. Phys.},
	author = {Schnoerr, David and Sanguinetti, Guido and Grima, Ramon},
	urldate = {2019-06-04},
	date = {2015-11},
	langid = {english}
}

@article{cao_linear_2018,
	title = {Linear mapping approximation of gene regulatory networks with stochastic dynamics},
	volume = {9},
	issn = {2041-1723},
	url = {http://www.nature.com/articles/s41467-018-05822-0},
	doi = {10.1038/s41467-018-05822-0},
	pages = {3305},
	number = {1},
	journaltitle = {Nat. Commun.},
	author = {Cao, Zhixing and Grima, Ramon},
	urldate = {2019-06-04},
	date = {2018-12},
	langid = {english}
}

@article{mcadams_its_1999,
	title = {It’s a noisy business! {G}enetic regulation at the nanomolar scale},
	volume = {15},
	issn = {0168-9525},
	url = {http://www.sciencedirect.com/science/article/pii/S016895259801659X},
	doi = {10.1016/S0168-9525(98)01659-X},
	pages = {65--69},
	number = {2},
	journaltitle = {Trends in Genetics},
	author = {{McAdams}, Harley H and Arkin, Adam},
	urldate = {2019-06-04},
	date = {1999-02}
}

@article{kiviet_stochasticity_2014,
	title = {Stochasticity of metabolism and growth at the single-cell level},
	volume = {514},
	issn = {0028-0836, 1476-4687},
	url = {http://www.nature.com/articles/nature13582},
	doi = {10.1038/nature13582},
	pages = {376--379},
	number = {7522},
	journaltitle = {Nature},
	author = {Kiviet, Daniel J. and Nghe, Philippe and Walker, Noreen and Boulineau, Sarah and Sunderlikova, Vanda and Tans, Sander J.},
	urldate = {2019-06-04},
	date = {2014-10},
	langid = {english}
}

@article{elowitz_stochastic_2002,
	title = {Stochastic gene expression in a single cell},
	volume = {297},
	issn = {00368075, 10959203},
	url = {http://www.sciencemag.org/cgi/doi/10.1126/science.1070919},
	doi = {10.1126/science.1070919},
	pages = {1183--1186},
	number = {5584},
	journaltitle = {Science},
	author = {Elowitz, M. B.},
	urldate = {2019-06-04},
	date = {2002-08},
	langid = {english}
}

@book{villani_optimal_2009,
	location = {Berlin Heidelberg},
	title = {Optimal Transport: Old and New},
	isbn = {978-3-540-71049-3},
	url = {https://www.springer.com/gb/book/9783540710493},
	series = {Grundlehren der mathematischen Wissenschaften},
	shorttitle = {Optimal Transport},
	publisher = {Springer},
	author = {Villani, Cédric},
	urldate = {2019-02-26},
	date = {2009},
	langid = {english}
}

@article{cuturi_sinkhorn_2013,
	title = {Sinkhorn Distances: Lightspeed Computation of Optimal Transport},
	volume = {26},
	pages = {2292--2300},
	journaltitle = {Advances in Neural Information Processing Systems},
	author = {Cuturi, Marco},
	date = {2013-04-07},
	langid = {english}
}

@article{frohlich_inference_2016,
	title = {Inference for stochastic chemical kinetics using moment equations and {System Size Expansion}},
	volume = {12},
	issn = {1553-7358},
	url = {https://journals.plos.org/ploscompbiol/article?id=10.1371/journal.pcbi.1005030},
	doi = {10.1371/journal.pcbi.1005030},
	pages = {e1005030},
	number = {7},
	journaltitle = {{PLOS} Computational Biology},
	shortjournal = {{PLOS} Comput. Bio.},
	author = {Fröhlich, Fabian and Thomas, Philipp and Kazeroonian, Atefeh and Theis, Fabian J. and Grima, Ramon and Hasenauer, Jan},
	urldate = {2018-09-21},
	date = {2016-07},
	langid = {english}
}

@article{schilling_adaptive_2016,
	title = {Adaptive moment closure for parameter inference of biochemical reaction networks},
	volume = {149},
	issn = {0303-2647},
	url = {http://www.sciencedirect.com/science/article/pii/S0303264716301204},
	doi = {10.1016/j.biosystems.2016.07.005},
	series = {Selected papers from the Computational Methods in Systems Biology 2015 conference},
	pages = {15--25},
	journaltitle = {Biosystems},
	author = {Schilling, Christian and Bogomolov, Sergiy and Henzinger, Thomas A. and Podelski, Andreas and Ruess, Jakob},
	urldate = {2018-09-21},
	date = {2016-11}
}

@article{zechner_moment-based_2012,
	title = {Moment-based inference predicts bimodality in transient gene expression},
	volume = {109},
	issn = {0027-8424, 1091-6490},
	url = {http://www.pnas.org/content/109/21/8340},
	doi = {10.1073/pnas.1200161109},
	pages = {8340--8345},
	number = {21},
	journaltitle = {Proceedings of the National Academy of Sciences},
	shortjournal = {{PNAS}},
	author = {Zechner, Christoph and Ruess, Jakob and Krenn, Peter and Pelet, Serge and Peter, Matthias and Lygeros, John and Koeppl, Heinz},
	urldate = {2018-09-21},
	date = {2012-05},
	langid = {english},
	pmid = {22566653}
}

@article{schnoerr_validity_2014,
	title = {Validity conditions for moment closure approximations in stochastic chemical kinetics},
	volume = {141},
	issn = {0021-9606, 1089-7690},
	url = {http://aip.scitation.org/doi/10.1063/1.4892838},
	doi = {10.1063/1.4892838},
	pages = {084103},
	number = {8},
	journaltitle = {The Journal of Chemical Physics},
	shortjournal = {J. Chem. Phys.},
	author = {Schnoerr, David and Sanguinetti, Guido and Grima, Ramon},
	urldate = {2018-10-25},
	date = {2014-08},
	langid = {english}
}

@article{choi_stochastic_2008,
	title = {A stochastic single-molecule event triggers phenotype switching of a bacterial cell},
	volume = {322},
	rights = {American Association for the Advancement of Science},
	issn = {0036-8075, 1095-9203},
	url = {http://science.sciencemag.org/content/322/5900/442},
	doi = {10.1126/science.1161427},
	pages = {442--446},
	number = {5900},
	journaltitle = {Science},
	author = {Choi, Paul J. and Cai, Long and Frieda, Kirsten and Xie, X. Sunney},
	urldate = {2018-11-27},
	date = {2008-10},
	langid = {english},
	pmid = {18927393}
}

@article{schnoerr_approximation_2017,
	title = {Approximation and inference methods for stochastic biochemical kinetics - a tutorial review},
	volume = {50},
	issn = {1751-8113, 1751-8121},
	url = {http://arxiv.org/abs/1608.06582},
	doi = {10.1088/1751-8121/aa54d9},
	pages = {093001},
	number = {9},
	journaltitle = {Journal of Physics A: Mathematical and Theoretical},
	shortjournal = {J. Phys. A},
	author = {Schnoerr, David and Sanguinetti, Guido and Grima, Ramon},
	urldate = {2018-12-03},
	date = {2017-03}
}

@article{marguerat_quantitative_2012,
	title = {Quantitative analysis of fission yeast transcriptomes and proteomes in proliferating and quiescent cells},
	volume = {151},
	issn = {0092-8674},
	url = {http://www.sciencedirect.com/science/article/pii/S0092867412011269},
	doi = {10.1016/j.cell.2012.09.019},
	pages = {671--683},
	number = {3},
	journaltitle = {Cell},
	shortjournal = {Cell},
	author = {Marguerat, Samuel and Schmidt, Alexander and Codlin, Sandra and Chen, Wei and Aebersold, Ruedi and Bähler, Jürg},
	urldate = {2019-06-10},
	date = {2012-10-26}
}

@article{shahriari_taking_2016,
	title = {Taking the human out of the loop: a review of {Bayesian} optimization},
	volume = {104},
	issn = {0018-9219, 1558-2256},
	url = {https://ieeexplore.ieee.org/document/7352306/},
	doi = {10.1109/JPROC.2015.2494218},
	shorttitle = {Taking the Human Out of the Loop},
	pages = {148--175},
	number = {1},
	journaltitle = {Proceedings of the {IEEE}},
	shortjournal = {Proc. {IEEE}},
	author = {Shahriari, Bobak and Swersky, Kevin and Wang, Ziyu and Adams, Ryan P. and de Freitas, Nando},
	urldate = {2019-06-21},
	date = {2016-01},
	langid = {english}
}

@article{thibault_overrelaxed_2017,
	title = {Overrelaxed {Sinkhorn-Knopp} algorithm for regularized Optimal Transport},
	url = {http://arxiv.org/abs/1711.01851},
	author = {Thibault, Alexis and Chizat, Lénaïc and Dossal, Charles and Papadakis, Nicolas},
	urldate = {2019-06-24},
	date = {2017-11-06},
	langid = {english},
	eprinttype = {arxiv},
	eprint = {1711.01851}
}

@book{rasmussen_gaussian_2006,
	location = {Cambridge, Mass},
	title = {Gaussian processes for machine learning},
	isbn = {978-0-262-18253-9},
	series = {Adaptive computation and machine learning},
	publisher = {{MIT} Press},
	author = {Rasmussen, Carl Edward and Williams, Christopher K. I.},
	date = {2006},
	langid = {english}
}

@article{friedman_linking_2006,
	title = {Linking stochastic dynamics to population distribution: an analytical framework of gene expression},
	volume = {97},
	issn = {0031-9007},
	doi = {10.1103/PhysRevLett.97.168302},
	shorttitle = {Linking stochastic dynamics to population distribution},
	pages = {168302},
	number = {16},
	journaltitle = {Physical Review Letters},
	shortjournal = {Phys. Rev. Lett.},
	author = {Friedman, Nir and Cai, Long and Xie, X. Sunney},
	date = {2006-10-20},
	pmid = {17155441}
}

@article{grima_steady-state_2012,
	title = {Steady-state fluctuations of a genetic feedback loop: An exact solution},
	volume = {137},
	issn = {0021-9606},
	url = {https://aip.scitation.org/doi/full/10.1063/1.4736721},
	doi = {10.1063/1.4736721},
	shorttitle = {Steady-state fluctuations of a genetic feedback loop},
	pages = {035104},
	number = {3},
	journaltitle = {The Journal of Chemical Physics},
	shortjournal = {J. Chem. Phys.},
	author = {Grima, R. and Schmidt, D. R. and Newman, T. J.},
	urldate = {2019-07-07},
	date = {2012-07-20}
}

@article{singh_derivative_2007,
	title = {A derivative matching approach to moment closure for the stochastic logistic model},
	volume = {69},
	issn = {0092-8240},
	doi = {10.1007/s11538-007-9198-9},
	pages = {1909--1925},
	number = {6},
	journaltitle = {Bulletin of Mathematical Biology},
	shortjournal = {Bull. Math. Biol.},
	author = {Singh, Abhyudai and Hespanha, João Pedro},
	date = {2007-08},
	pmid = {17443391}
}

@article{soltani_conditional_2015,
	title = {Conditional Moment Closure Schemes for Studying Stochastic Dynamics of Genetic Circuits},
	volume = {9},
	issn = {1932-4545},
	doi = {10.1109/TBCAS.2015.2453158},
	pages = {518--526},
	number = {4},
	journaltitle = {{IEEE} Transactions on Biomedical Circuits and Systems},
	shortjournal = {{IEEE} Trans. Biomed. Circuits Syst.},
	author = {Soltani, M. and Vargas-Garcia, C. A. and Singh, A.},
	date = {2015-08}
}

@article{ramaswamy_discreteness-induced_2012,
	title = {Discreteness-induced concentration inversion in mesoscopic chemical systems},
	volume = {3},
	rights = {2012 Nature Publishing Group},
	issn = {2041-1723},
	url = {https://www.nature.com/articles/ncomms1775},
	doi = {10.1038/ncomms1775},
	pages = {779},
	journaltitle = {Nature Communications},
	shortjournal = {Nat. Commun.},
	author = {Ramaswamy, Rajesh and González-Segredo, Nélido and Sbalzarini, Ivo F. and Grima, Ramon},
	urldate = {2019-07-09},
	date = {2012-04-10},
	langid = {english}
}

@article{gutmann_bayesian_2016,
	title = {Bayesian Optimization for Likelihood-Free Inference of Simulator-Based Statistical Models},
	volume = {17},
	pages = {1--47},
	number = {125},
	journaltitle = {Journal of Machine Learning Research},
	shortjournal = {J. Mach. Learn. Res.},
	author = {Gutmann, Michael U. and Corander, Jukka},
	urldate = {2019-07-09},
	date = {2016},
	langid = {english}
}

@article{vazquez_convergence_2010,
	title = {Convergence properties of the expected improvement algorithm with fixed mean and covariance functions},
	volume = {140},
	issn = {0378-3758},
	url = {http://www.sciencedirect.com/science/article/pii/S0378375810001850},
	doi = {10.1016/j.jspi.2010.04.018},
	pages = {3088--3095},
	number = {11},
	journaltitle = {Journal of Statistical Planning and Inference},
	shortjournal = {J. Stat. Plan. Inference},
	author = {Vazquez, Emmanuel and Bect, Julien},
	urldate = {2019-07-09},
	date = {2010-11-01}
}

@article{snoek_input_2014,
	title = {Input Warping for {Bayesian} Optimization of Non-Stationary Functions},
	url = {http://proceedings.mlr.press/v32/snoek14.html},
	pages = {1674--1682},
	journal = {Proceedings of the International Conference on Machine Learning},
	volume = {31},
	author = {Snoek, Jasper and Swersky, Kevin and Zemel, Rich and Adams, Ryan},
	urldate = {2019-07-08},
	date = {2014-01-27},
	langid = {english}
}

@article{ashyraliyev_systems_2009,
	title = {Systems biology: parameter estimation for biochemical models},
	volume = {276},
	rights = {© 2009 The Authors Journal compilation © 2009 {FEBS}},
	issn = {1742-4658},
	url = {https://febs.onlinelibrary.wiley.com/doi/abs/10.1111/j.1742-4658.2008.06844.x},
	doi = {10.1111/j.1742-4658.2008.06844.x},
	shorttitle = {Systems biology},
	pages = {886--902},
	number = {4},
	journaltitle = {The {FEBS} Journal},
	author = {Ashyraliyev, Maksat and Fomekong‐Nanfack, Yves and Kaandorp, Jaap A. and Blom, Joke G.},
	urldate = {2019-10-14},
	date = {2009},
	langid = {english}
}

\clearpage 

\begin{landscape}

\appendix

\bigbreak

\section{Results for moment-based inference}

\label{sec:apdx_results}

\vfill

\begin{table}[H]
\centering
\footnotesize
\begin{tabu}{| c | c | c | c | c | c | c | c | c | c | c | c | c | c | c | c |}
\tabucline{1-16}
\multirow{2}{*}{\textbf{Method}} & \multicolumn{3}{c|}{\textbf{PFL}} & \multicolumn{3}{c|}{\textbf{NFL}} & \multicolumn{3}{c|}{\textbf{R1}} & \multicolumn{3}{c|}{\textbf{R2}} & \multicolumn{3}{c|}{\textbf{R3}} \\
\tabucline{2-16}
& $\bm{\rho_u}$ & $ \bm{\sigma_u} $ & $\bm{\sigma_b}$& $\bm{\rho_u}$ & $ \bm{\sigma_u} $ & $\bm{\sigma_b}$& $\bm{\rho_u}$ & $ \bm{\sigma_u} $ & $\bm{\sigma_b}$& $\bm{\rho_u}$ & $ \bm{\sigma_u} $ & $\bm{\sigma_b}$& $\bm{\rho_u}$ & $ \bm{\sigma_u} $ & $\bm{\sigma_b}$ \\
\tabucline{1-16}
GT & 0.3 & 400 & 2.5 & 13 & 0.1 & 0.001 & 1.3 & 42 & 1.5 & 9.3 & 1 & 0.7 & 1 & 355 & 7.9 \\
\tabucline{1-16}
CDM1 & 0.15 & 1000.0 & 5.82 & 12.6 & 0.025 & 0.0002 & 43.2 & 100.0 & 0.10 & 3.26 & 1.62 & 0.26 & 0.10 & 356.5 & 10.0 \\
CDM2 & 0.97 & 74.7 & 5.13 & 2.14 & 0.031 & 0.0076 & 1.20 & 54.1 & 0.71 & 1.65 & 0.19 & 0.11 & 3.73 & 12.1 & 2.22 \\
CG1 & 1.41 & 1000.0 & 5.45 & 12.6 & 0.012 & 0.0001 & 43.2 & 100.0 & 0.10 & 3.32 & 0.66 & 0.11 & 1.36 & 326.3 & 6.67 \\
CG2 & 7.78 & 619.6 & 2.07 & 12.9 & 0.070 & 0.0007 & 40.2 & 50.2 & 0.17 & 1.49 & 10.0 & 0.10 & 2.46 & 192.7 & 2.38 \\
CNB1 & 0.15 & 1000.0 & 5.86 & 13.9 & 0.24 & 0.0032 & 43.2 & 100.0 & 0.10 & 4.21 & 1.25 & 0.30 & 0.10& 341.3 & 10.0 \\
CNB2 & 0.42 & 39.1 & 0.98 & 86.2 & 0.051 & 0.0027 & 4.38 & 31.3 & 7.91 & 3.45 & 1.27 & 7.77 & 0.62 & 552.5 & 0.30 \\
Gauss & - & - & - & 12.6 & 0.017 & 0.0002 & 43.2 & 100.0 & 0.10 & - & - & - & - & - & -  \\
DM & 0.15 & 1000.0 & 5.90 & 14.3 & 0.29 & 0.0042 & 43.2 & 100.0 & 0.10 & 3.74 & 1.23 & 0.25 & 0.10 & 336.1 & 10.0 \\
LMA & 15.0 & 1000.0 & 0.79 & 13.7 & 0.068 & 0.0008 & 43.2 & 100 & 0.10 & 6.75 & 0.75 & 0.19 & 4.53 & 1000.0 & 0.10 \\
\tabucline{1-16}
\end{tabu}

\captionsetup{width=0.9\textwidth}
\caption{Ground truth and inferred parameters for the five reaction networks considered in \cref{sec:exp} using the tested moment-closure schemes. Gaussian moment closure failed to converge in most cases.}

\end{table}

\vfill

\begin{table}[H]
\footnotesize
\centering
\begin{tabu}{| c | c | c | c | c | c | c | c | c | c | c |}
\tabucline{1-11}
\multirow{2}{*}{\textbf{Method}} & \multicolumn{2}{c|}{\textbf{PFL}} & \multicolumn{2}{c|}{\textbf{NFL}} & \multicolumn{2}{c|}{\textbf{R1}} & \multicolumn{2}{c|}{\textbf{R2}} & \multicolumn{2}{c|}{\textbf{R3}} \\
\tabucline{2-11}
& $\bm{m_P}$ & $ \bm{s_P} $ & $\bm{m_P}$ & $ \bm{s_P}$ & $\bm{m_P}$ & $ \bm{s_P}$ & $\bm{m_P}$ & $ \bm{s_P} $ & $\bm{m_P}$ & $ \bm{s_P} $ \\
\tabucline{1-11}
GT & 32.6 & 28.0 & 27.9 & 19.5 & 66.4 & 16.1 & 11.4 & 10.1 & 9.9 & 9.3 \\
\tabucline{1-11}
CDM1 & 15.8 & 22.8 & 28.4 & 19.2 & 66.5 & 13.1 & 11.8 & 9.8 & 3.5 & 7.6 \\
CDM2 & 195 & 25.2 & 2.57 & 4.39 & 16.6 & 15.1 & 5.3 & 7.1 & 43.6 & 12.9\\
CG1 & 31.9 & 21.4 & 28.7 & 19.1 & 66.6 & 13.1 & 11.1 & 10.4 & 10.7 & 9.0  \\
CG2 & 36.6 & 15.2 & 28.5 & 19.3 & 66.5 & 13.5 & 11.6 & 10.1 & 10.7 & 7.8 \\
CNB1 & 13.2 & 20.7 & 27.8 & 19.6 & 66.7 & 13.3 & 11.8 & 10.2 & 3.7 & 7.9 \\
CNB2 & 169 & 26 & 25.8 & 61.9 & 90.5 & 15.5 & 4.8 & 6.2 & 1.5 & 2.3 \\
Gauss & - & - & 29.4 & 19.0 & 66.4 & 13.1 & - & - & - & - \\
DM & 16.8 & 23.1 & 27.9 & 19.7 & 66.7 & 13.1 & 11.8 & 10.0 & 4.2 & 8.5\\
LMA & 34.8 & 11 & 27.8 & 21.0 & 67.1 & 13.1 & 13.7 & 12.2 & 11.0 & 6.2 \\
\tabucline{1-11}
\end{tabu}

\captionsetup{width=0.65\textwidth}
\caption{Mean and standard deviation over protein counts for the ground truth parameters and those inferred using moment-based inference.}

\end{table}

\vfill

\end{landscape}

\end{document}